\documentclass[%
 aip,
 amsmath,amssymb,
 reprint,%
 nofootinbib, rmp
]{revtex4-1}

\usepackage{graphicx}
\usepackage{dcolumn}
\usepackage{bm}

\usepackage[utf8]{inputenc}
\usepackage[T1]{fontenc}
\usepackage{etoolbox}

\usepackage{graphics}

\usepackage{graphicx}
\usepackage{upgreek}
\usepackage{isomath}

\usepackage{color}
\usepackage{xcolor}

\definecolor{green}{rgb}{0.0, 0.666, 0.0}

\definecolor{darkgreen}{rgb}{0.55, 0.71, 0.0}

\definecolor{lightviolet}{rgb}{0.79, 0.28, 0.96}

\definecolor{navy}{rgb}{0.05, 0.40, 0.8}

\definecolor{purple}{rgb}{0.62, 0.0, 0.666}

\usepackage{amsmath, amssymb, amsfonts, amstext}
\usepackage{pifont}
\usepackage{mathrsfs}

\colorlet{linkequation}{blue}
\usepackage[hidelinks]{hyperref}
\hypersetup{
    colorlinks = true,
    citecolor = blue,
    linkcolor=blue
}

\makeatletter
\def\@email#1#2{%
 \endgroup
 \patchcmd{\titleblock@produce}
  {\frontmatter@RRAPformat}
  {\frontmatter@RRAPformat{\produce@RRAP{*#1\href{mailto:#2}{#2}}}\frontmatter@RRAPformat}
  {}{}
}%
\makeatother
\begin{document}

\preprint{arXiv:2302.12819}

\title{On the origin of the negative energy-related contribution to the elastic modulus of rubber-like gels}
\author{L. K. R. Duarte}
\affiliation{ 
Departamento de F\'isica, Universidade Federal de Vi\c{c}osa (UFV), Av. P. H. Rolfs, s/n, 36570-900, Vi\c{c}osa, Brazil
}%
\affiliation{ 
Instituto Federal de Educa\c{c}\~ao, Ci\^encia e Tecnologia de Minas Gerais, P\c{c}. Jos\'e Emiliano Dias, 87, 35430-034, Pte. Nova, Brazil
}%
\author{L. G. Rizzi}%
\affiliation{ 
Departamento de F\'isica, Universidade Federal de Vi\c{c}osa (UFV), Av. P. H. Rolfs, s/n, 36570-900, Vi\c{c}osa, Brazil
}%

\date{\today}

\begin{abstract}
	We consider a coarse-grained polymer model in order to investigate the origin of a recently discovered negative energy-related contribution to the elastic modulus $G(T)$ of rubber-like gels.
	From this model, we are able to compute an exact expression for the free energy of the system, which allows us to evaluate a stress-strain relationship that displays a non-trivial dependence on the temperature $T$.
	We validate our approach through comparisons between the theoretical results and the experimental data obtained for tetra-PEG hydrogels, which indicate that, although simple, the present model works well to describe the experiments.
	Importantly, our approach unveiled aspects of the experimental analysis which turned out to be different from the conventional entropic and energetic analysis broadly used in the literature. Also,
	in contrast to the linear dependence predicted by the traditional, {\it i.e.}, purely entropic, models, our results suggest that the general expression of the elastic modulus should be of the form $G(T) \propto k_BT w(T)$, with $w(T)$ being a temperature-dependent correction factor that could be related to the interaction between the chains in the network and the solvent.
	Accordingly, the correction factor allows the expression found for the elastic modulus to describe both rubber and rubber-like gels.
\end{abstract}

\maketitle

\section{Introduction}
\label{intro}

	In contrast to many solid materials, rubber-like gels are semisolid viscoelastic materials formed by networks of disordered crosslinked polymeric chains immersed in liquid solvents which display isotropic and soft mechanical properties~\cite{richtering2014softmatter}.
	Examples~\cite{annabi2014advmater,thiele2014advmater,kamata2015advhmater,calo2015eurpolj,basu2011macromol,wen2012softmatter,dai2021softmatter} include not only biologically relevant but also synthetically designed gels, {\it e.g.}, modified hyaluronic acids,  polyethylene glycol (PEG), and polyacrylamide (PAM) gels.
	Due to the possibility of having a wide range of elasticity, mechanical strength, porosity, and swellability, hydrogels are materials that hold great promise in several applications~\cite{gu2020angewchem,danielsen2021chemrev} such as, {\it e.g.}, drug delivery systems and tissue engineering scaffolds.

	Remarkably, because of the incompressibility that is generally observed in experiments~\cite{yoshikawa2021prx}, the mechanical response of rubber-like gels is usually characterized by a single elastic constant\footnote{For rubber-like gels, $G(T)$ corresponds to the shear modulus, which is usually much smaller than the bulk modulus $B(T)$, even so, the former governs their response to locally applied stresses and it also determines changes in their shape.} $G(T)$ (see the discussions in Ref.~\cite{treloarbook}).
	It is called the 
elastic modulus 
and it is usually assumed to be given by~\cite{treloarbook,toda2018aipadv}
\begin{equation}
G(T) = n_e k_BT~~,
\label{GTgeneral}
\end{equation}
where $k_B$ is the Boltzmann's constant, $T$ is the absolute temperature of the solvent, and $n_e$ is the effective number density of ``elastic elements'' in solution.
	 Different models~\cite{treloarbook,sakumichi2021polymj} have different interpretations for $n_e$, but virtually all theoretical approaches assume that it should be proportional to the number density of precursor molecules in solution, $n$, since both the number density of strands ({\it i.e.}, bridged chains), $n_f$, and the number density of crosslinks, $n_c$, should be proportional to $n$.

	Interestingly, most of the theoretical approaches agreed on the linear dependence of $G(T)$ on the temperature $T$ given by Eq.~\ref{GTgeneral}.
	Essentially, this occurs because energetic contributions were neglected in the development of the seemingly successful models~\cite{treloarbook,sakumichi2021polymj}, {\it e.g.,} those that are based on the classical affine~\cite{flory1953book}, phantom network~\cite{james1953jcp}, and junction affine~\cite{flory1977jcp} models.
	In other words, those models consider that only entropic effects of the chains in the system are responsible to the observed temperature-dependent elastic modulus.
	It seems that such approximation works fine for natural and synthetic rubber~\cite{treloarbook,anthony1942jpc},
but when a purely entropic approach is used to describe rubber-like gels, inconsistencies have been found between the theoretical predictions and the experimental results.

	Indeed, recent experiments have shown that the linear elasticity of some gels can be fully described only when a significant negative 
contribution to its elastic modulus is considered~\cite{yoshikawa2021prx,fujiyabu2021prl}. 
	This issue was analyzed experimentally for tetra-PEG hydrogels that form networks with well-controlled topological structures~\cite{yoshikawa2021prx,sakumichi2021polymj}, and a phenomenological correction to the linear description was proposed in order to consider such a negative 
contribution.
	In particular, instead of assuming $G(T)$ to be proportional to the absolute temperature $T$, as in Eq.~\ref{GTgeneral}, the authors in Ref.~\cite{yoshikawa2021prx} added an {\it ad hoc} term to the elastic modulus, writing it as
\begin{equation}
G(T)=a(T-T_0)~~,
\label{heurG}
\end{equation}
where $a>0$ and $T_0>0$ are parameters that depend on the experimental conditions that were used to create the gels such as the concentration of precursor molecules in solution ($c$), their molar mass ($M$), and the connectivity ($p$) of the network~\cite{yoshikawa2021prx,sakumichi2021polymj}.
	Since the first term in this equation is proportional to $T$, {\it i.e.}, $G_S(T) = aT$, the authors of Ref.~\cite{yoshikawa2021prx} termed it as an {\it entropic contribution} to the elastic modulus, whereas the second one, $G_E(T) = -aT_0$, is its negative energy-related contribution, so that the total elastic modulus is given by $G(T) = G_S(T) + G_E(T)$. 

	Even though the authors have used Eq.~\ref{heurG} to describe the experimental results presented in Ref.~\cite{yoshikawa2021prx}, its origin is still an open issue.
	In our view, the qualitative microscopic model discussed in Ref.~\cite{yoshikawa2021prx} correctly presumes that the interaction between the chains in the network and their neighbouring solvent molecules should play an important role in the viscoelastic behaviour of rubber-like gels as the tetra-PEG hydrogels.
	However, we believe that the energy $E_s$ of the chain in its elongated ({\it e.g.}, swelled, {\it s}) conformation should not be smaller than the energy $E_b$ of its collapsed ({\it e.g.}, blob, {\it b}) conformation, that is, $E_b > E_s$, as assumed in the model discussed in Ref.~\cite{yoshikawa2021prx}.
	It is worth noting that this feature was also included as an effective self-repulsive interaction in the statistical mechanical model presented in Ref.~\cite{shirai2022arxiv} that attempts to explain the origin of the negative contribution to $G(T)$.
	Nonetheless, such qualitative behaviour contrasts with results observed from numerical simulations on folding phenomena~\cite{junghans2006prl,chen2008pre,liu2012jcp,frigori2013jcp}, where $E_b < E_s$, {\it i.e.,} chains in their elongated states have higher energies than the collapsed ones.
	Indeed, as it can be argued from the results presented latter in our work, one should have $E_s > E_b$ in order to explain the origin of the negative contribution $G_E(T)$ in rubber-like gels.

	Accordingly, here we consider one of the simplest mesoscopic models which could be explored to explain the negative contribution $G_E(T)$ to the elastic modulus $G(T)$ of rubber-like hydrogels.
	In addition, such a simple model might provides us an interpretation for the phenomenological expression given by Eq.~\ref{heurG}.
	In fact, our results indicate that it is possible to obtain expressions for the elastic modulus for these kind of 
gels in the form
\begin{equation}
G(T) \propto k_BT w(T)~~,
\label{theorG}
\end{equation}
where $w(T)$ is a correction factor that makes the classical models able to describe not only purely entropic rubbers but rubber-like gels as well.
	This result is validated for the whole set of experimental data presented in Refs.~\cite{yoshikawa2021prx}, where tetra-PEG hydrogels were studied.
 
	The remainder of the manuscript is as follows.
	In Sec.~\ref{model} we present the model and discuss how a force-extension relationship can be obtained from its thermostatistics.
	Most of the results related to the stress-strain relationship and to the elastic moduli of the model, as well as their experimental validation, are presented in Sec.~\ref{results}.
	Also, in Sec.~\ref{results} we discuss how the theoretical results obtained from this simple model might be related to the phenomenological expression given by Eq.~\ref{heurG}.
	Finally, our main conclusions are presented in Sec.~\ref{conclusions}.

\section{Model and its thermostatistics}
\label{model}

\begin{figure}[!t]
\centering
\resizebox{0.45\textwidth}{!}{%
\includegraphics{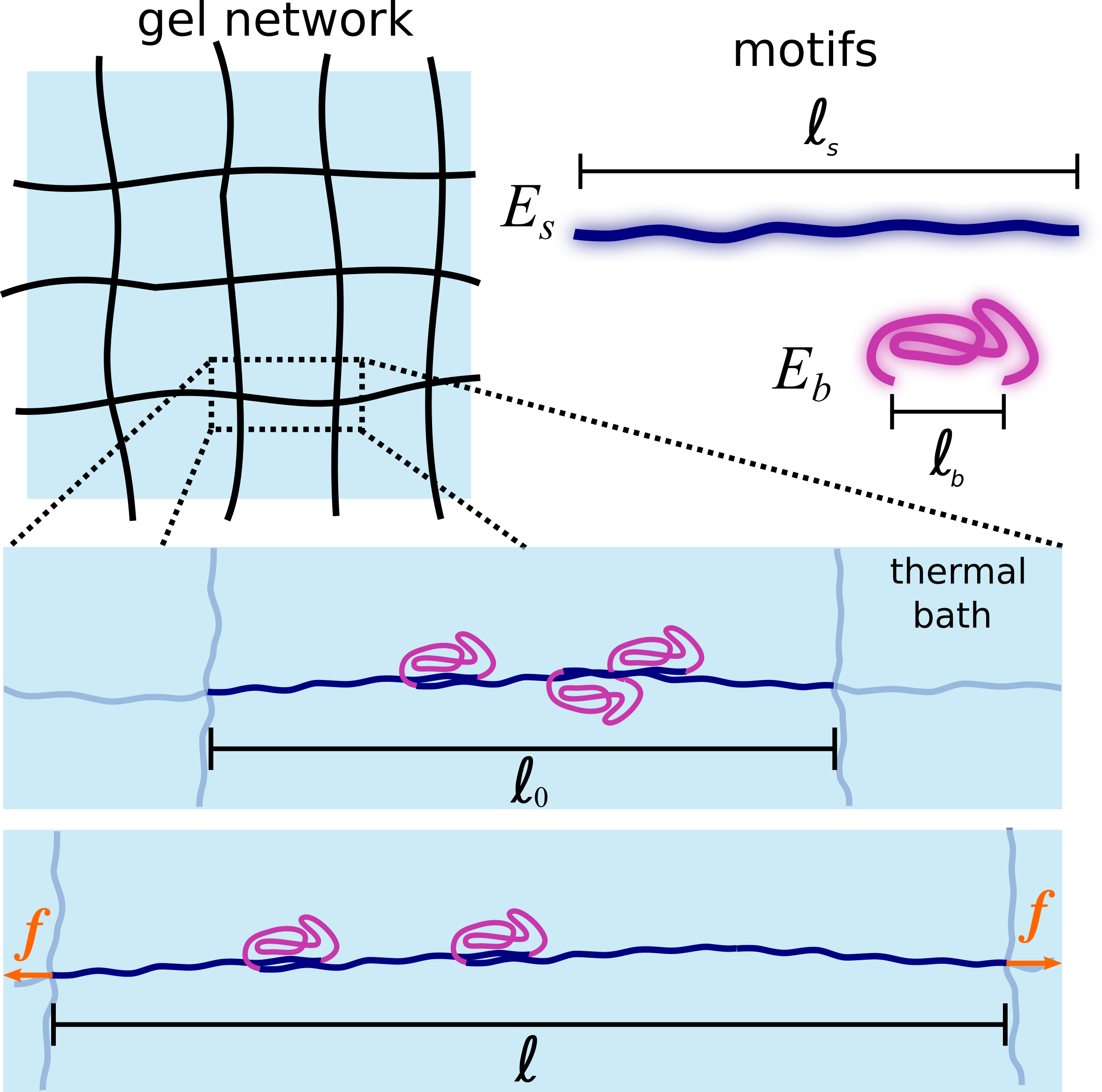}
}
\caption{The chains in the gel network are described by a simple coarse-grained model with $N$ segments that can be in one of two possible conformational motifs: the elongated, or swelled ({\it s}), state, with energy $E_s$ and length $\ell_s$, and the collapsed, or blob ({\it b}), state, which has energy $E_b$ and length $\ell_b$.
	The lengths $\ell_s$ and $\ell_b$ can be viewed as steps of an one-dimensional random-walk, where the right step is positive and the left step is negative.
	The natural ({\it i.e.}, relaxed) and the stretched ($f \neq 0$) end-to-end distances of the chain are denoted by $\ell_0$ and $\ell$, respectively.}
\label{fig:model}
\end{figure}

	For simplicity, we consider a coarse-grained model where a bridged chain is assumed to have $N$ segments that can be only in one of two possible conformational motifs: the elongated, or swelled ({\it s}), and the collapsed, or blob ({\it b}), states.
	As showed in Fig.~\ref{fig:model}, the swelled state corresponds to a longer conformation with length $\ell_s$ and energy $E_s$, while the blob state is characterized by a conformation with length $\ell_b$ and energy $E_b < E_s$, so that $\Delta \ell = \ell_s - \ell_b >0$  and $\Delta E = E_s - E_b>0$.
	As mentioned before, the assumption that $E_s > E_b$ is supported by numerical simulations performed with effective coarse-grained models that assume implicity interactions with the solvent to describe the collapse of the chain (see, {\it e.g.}, Refs.~\cite{junghans2006prl,chen2008pre,liu2012jcp,frigori2013jcp}).
	The negative value\footnote{Actually, the negative value of $\ell_b$ is not an issue if one thinks about this model as representing a one-dimensional random walk with different energy costs for the backward and the forward steps of different sizes. In fact, the one-dimensional freely-jointed chain (FJC) model~\cite{kubobook} is a special case of the model presented here when one imposes $E_s = E_b$ and $\ell_s = - \ell_b$.} 
	of $\ell_b$ is explained in Fig.~\ref{fig:model}, and, as we will show later, it is a necessary feature for the model in order to have the possibility of having zero end-to-end distances, {\it i.e.}, $\ell=0$.
	This is interesting because it seems to be the main qualitative behaviour observed for Gaussian chains that have been used to describe rubber-like gels~\cite{nishi2015jcp}.
	It is worth noting that, for simplicity, we assume that the given energies, $E_s$ and $E_b$, and lengths, $\ell_s$ and $\ell_b$, of the individual segments in the two conformational states do not depend neither on the force $f$ or on the temperature $T$ of the thermal bath ({\it i.e.}, the portion of the solvent which does not interact directly with the chain).
	Obviously, the end-to-end distance of the chain
 will depend on both thermodynamic variables.

	Now, by considering that the chain has $N=n_s + n_b$ segments, with $n_s$ and $n_b$ being the number of segments in the swelled and in the blob states, respectively, the number of possible configurations of the system is given by
\begin{equation}
\mathrm{\Omega}(N,n_b,n_s) = \frac{N!}{n_b! \, n_s!}~~.
\label{dos-nsnb} 
\end{equation}
	Thus, the end-to-end distance of the chain can be written as
\begin{equation}
\ell(n_s, n_b) = n_s \ell_s + n_b \ell_b~~,
\label{end-to-end-nsnb}
\end{equation}
and its internal energy, which includes the interaction between the chain and their neighboring solvent molecules, is given by
\begin{equation}
U(n_s, n_b) = n_s E_s + n_b E_b ~~.
\label{internalenergy-nsnb}
\end{equation}

	The thermostatistics of such simple model can be determined either on the {\it constant-stress ensemble}, where the force $f$ is fixed and the end-to-end distance $\ell$ can fluctuate, or on the {\it constant-strain ensemble}, where $\ell$ is fixed and $f$ can vary. 
	Since both approaches lead to exactly the same force-extension relation, in the following we will present the results in terms of the \textit{constant-stress ensemble}, while in the Appendix~\ref{constant-strain} we demonstrate how the same relation can be obtained from the \textit{constant-strain ensemble}.

\subsection{Relationship between force and extension}

	By considering that the chain is subjected to a constant force $f$ at its ends, and by suppressing the dependence on $n_s$ in Eqs.~\ref{dos-nsnb},~\ref{end-to-end-nsnb}, and~\ref{internalenergy-nsnb}, since $n_s=N-n_b$, one can write the probability density function related to finding a specific state of the system as
\begin{equation}
p(N,n_b) = \frac{\mathrm{\Omega}(N,n_b) \, 
e^{-H(N,n_b)/k_BT }}{\mathcal{Z}_{\mathcal{G}}(N,T,f)}~~,
\end{equation}
where the function $H(N,n_b)=U(N,n_b) - f\ell(N,n_b)$ can be thought as an enthalpy.
	The normalization constant corresponds to the  partition function of the system and is given by 
$\mathcal{Z}_{\mathcal{G}}(N,T,f) = \sum_{n_b=0}^{N} \mathrm{\Omega}(N,n_b) \, e^{-H(N,n_b)/k_BT }$, 
which yields~\cite{kubobook}
\begin{equation}
\mathcal{Z}_{\mathcal{G}}(N,T,f) = \left[ e^{(f \ell_s - E_s)/k_BT} + e^{(f \ell_b - E_b)/k_BT} \right]^{N}~~.
\label{Yresult}
\end{equation}

	Hence, the above result can be used to compute the following free energy ({\it i.e.}, Gibbs thermodynamic potential) $\mathcal{G}(N,T,f) \equiv - k_B T \, \ln[ \mathcal{Z}_{\mathcal{G}}(N,T,f) ]$, 
which leads to
\begin{equation}
\mathcal{G}(N,T,f) =-Nk_BT\ln\left[ e^{(f \ell_s - E_s)/k_BT} + e^{(f \ell_b - E_b)/k_BT} \right]~~,
\label{Gibbs}
\end{equation}
from where one can obtain all the thermostatistical properties of the system.
	In particular, one can estimate the average value of the end-to-end distance of the chain as\footnote{For the free energy defined by the potential $\mathcal{G}$ one has the differential form $d\mathcal{G}= - SdT - \ell df$ where $S$ and $\ell$ are the entropy and the end-to-end distance of the system, respectively.}
\begin{equation}
\ell
= - \left[ \frac{ \partial }{\partial f} \mathcal{G}(N,T,f) \right]_{N,T}
= N \left[ \frac{\ell_s  + \ell_b  e^{(\Delta E - f \Delta \ell)/k_BT}   }{ 1 +  e^{(\Delta E - f \Delta \ell)/k_BT} } \right]~.
\label{end-to-end}
\end{equation}

	It is worth noting that, at finite temperatures, this result indicates that one may have a non-zero end-to-end distance even in the absence of forces.
	So, we use Eq.~\ref{end-to-end} with $f=0$ to define the natural ({\it i.e.}, relaxed) end-to-end distance as $\ell_0 \equiv \ell(N,T,0)$, that is,
\begin{equation}
\ell_0  = N \left(  \frac{\ell_s  + \ell_b  e^{\Delta E/k_BT}   }{ 1 +  e^{\Delta E/k_BT} } \right)~~.
\label{ell0}
\end{equation}
	Thus, Eqs.~\ref{end-to-end} and~\ref{ell0} can be used to define the strain $\gamma \equiv \gamma(T,\ell)$ through the following relation
\begin{equation}
\gamma = \frac{\ell - \ell_0}{\ell_0}~~.
\label{gamma-strain}
\end{equation}
	We note that, 
since $\ell \equiv \ell(N,T,f)$ and $\ell_0 \equiv \ell(N,T,0)$,
the strain $\gamma$ is a function of the temperature and the force applied to the chain. 
	As expected, this definition is consistent with having $\gamma=0$ if $f = 0$ for any value of the temperature $T$.
	Also, expression~\ref{gamma-strain} indicates that the strain $\gamma$ can be either positive or negative, as it only depends whether the chain is being stretched ($\ell > \ell_0$) or compressed ($\ell < \ell_0$).

	In the \textit{constant-stress ensemble}, one can evaluate the force in terms of the end-to-end distance by inverting Eq.~\ref{end-to-end}, which yields the seemingly 
 known result given by~\cite{kubobook}
\begin{equation}
f(T,\ell) = \frac{\Delta E}{\Delta \ell} + \frac{k_BT}{\Delta \ell} \ln \left( \frac{\ell - N\ell_b}{N \ell_s - \ell}  \right)~~.
\label{force-dist}
\end{equation} 
	It is worth noting that, as we mentioned before, here we assume that $\ell_b<0$, so that $\ell$ corresponds to the end-to-end distance and not the total ({\it i.e.}, contour) length of the chain. 
	Besides its different interpretation, we indicate that, in that case, Eq.~\ref{force-dist} is physically well-defined only for $0 \,\leq\, \ell < N \ell_s$. 
	In addition, since we assume that $\ell_b < 0$ and require that the natural end-to-end distance of the chain $\ell_0$ to be positive by definition, one has to consider that Eqs.~\ref{ell0}, \ref{gamma-strain}, and \ref{force-dist} will have a precise physical meaning only for temperatures $T$ higher than a characteristic temperature $T_0^{*}$, which is found from the condition that $\ell(N,T_0^*,0) = 0$. 
	The consequences of this condition are better discussed in the Sec.~\ref{Tzero}.

\section{Results}
\label{results}

	In this section we discuss how the results obtained for the present model can be used to provide a suitable framework to describe the experimental data extracted from Ref.~\cite{yoshikawa2021prx} for tetra-PEG hydrogels.
	In particular, we will focus on the mechanical response of the gel networks described by stress-strain curves and also on the temperature-dependent behaviour of the elastic modulus.

\subsection{$(T,\gamma)$-dependent stress}
\label{Tgamma-depend-sigma}

	Accordingly, in order to link the force-extension behaviour of a single chain to the response of the whole system, we assume that the force should be related to the stress as $f = A \sigma$, where $A$ is an effective cross-sectional area\footnote{Here, the effective cross-sectional area $A$ related to the volume occupied by a bridged chain is assumed to be independent of the temperature $T$, however, as it will be argued later, it might depend on other quantities such as the concentration of precursor molecules ($c$) in solution, the molar mass ($M$) of such molecules, and the connectivity ($p$) of the network.
	Also, since we are dealing with an implicit averaging procedure that should be done over different orientations of the chains of a disordered isotropic system, possible differences related to the way one applies the stress and, consequently, to the way one defines the elastic modulus, are immaterial and should be incorporated into $A$.}
	perpendicular to the direction of the force. 
	Hence, by considering Eqs.~\ref{gamma-strain} and~\ref{force-dist} one can write the temperature-dependent stress in terms of the strain as\footnote{As shown in Appendix~\ref{Helmholtzfreenergy}, this result can be also obtained directly from the derivative of the Helmholtz free energy density and it corresponds to an {\it engineering stress}~\cite{treloarbook}.}
\begin{equation}
\sigma(T,\gamma) = \frac{\Delta E}{A\Delta \ell} + \frac{k_BT}{A\Delta \ell} \ln \left(
\frac{\gamma_b + \gamma}{ \gamma_s - \gamma }
\right)~~,
\label{sigma-gamma}
\end{equation}
where 
\begin{equation}
\gamma_s = \frac{N \ell_s - \ell_0}{\ell_0}~~,
\label{gammas}
\end{equation}
and
\begin{equation}
\gamma_b = \frac{\ell_0 - N\ell_b}{\ell_0}~~,
\label{gammabb}
\end{equation}
with $\ell_0$ given by Eq.~\ref{ell0}.

	We note that the above definitions of $\gamma_s$ and $\gamma_b$ are consistent with $\sigma(T,0) = 0$, thus, from Eq.~\ref{sigma-gamma}, one finds that
\begin{equation}
\gamma_b = \gamma_s \, e^{-\Delta E/k_BT}~~,
\label{gammab}
\end{equation}
which can be also verified through Eqs.~\ref{gammas} and~\ref{gammabb} with the definition of $\ell_0$ given by Eq.~\ref{ell0}.
	Also, it is worth noting that $\gamma_s\equiv\gamma_s(T)$ and $\gamma_b\equiv\gamma_b(T)$ will both present a dependence on the temperature because $\ell_0\equiv \ell(N,T,0)$, however, since they are defined as ratios between lengths, the stress $\sigma(T,\gamma)$ will not depend on the value of the number of segments $N$.
	As we will see later, the stress-strain relation given by Eq.~\ref{sigma-gamma} leads $\sigma(T,\gamma)$ to display a non-linear behaviour as a function of the temperature.

	Interestingly, for small strains, one can assume that $\gamma/\gamma_s \ll 1$ and $\gamma/\gamma_b \ll 1$ (as a consequence of Eq.~\ref{gammab}). 
	Hence, by considering the linearization $\ln(1+x) \simeq x $, the stress given by Eq.~\ref{sigma-gamma} can be approximated by 
\begin{equation}
\sigma(T,\gamma) \simeq \frac{k_BT}{\gamma_sA\Delta \ell} 
 \left( 1 + e^{\Delta E/k_BT} \right) \gamma ~~.
\label{sigma-small-gamma}
\end{equation}

\subsection{$T$-dependent elastic modulus}

	 By considering the stress $\sigma(T,\gamma)$, one can evaluate the temperature-dependent elastic modulus as~\cite{yoshikawa2021prx}
\begin{equation}
G(T) = \lim_{\gamma \rightarrow 0} \left[ \frac{\partial }{\partial \gamma}\sigma(T,\gamma) \right]_{T} ~~. 
\label{GTdef}
\end{equation}
	Hence, using the result~\ref{sigma-gamma} one finds that
\begin{equation}
G(T) = \frac{k_BT}{\gamma_s A \Delta \ell} \left( 1 + e^{\Delta E/k_BT} \right)  ~~,
\label{elasticmodulus-total}
\end{equation}
which is exactly the same result that would be obtained if one had replaced the linearized expression for stress, Eq.~\ref{sigma-small-gamma}, in the definition of $G(T)$ given by Eq.~\ref{GTdef}.
	Accordingly, the above expression can be identified as the temperature-dependent pre-factor in Eq.~\ref{sigma-small-gamma}, which can be 
rewritten as
\begin{equation}
\sigma(T,\gamma) \simeq G(T) \gamma ~~,
\label{sigma-gamma-linear}
\end{equation}
demonstrating that, at least in the linear regime, the limit of taking a small value for $\gamma$ is interchangeable with the partial derivative in Eq.~\ref{GTdef}.

	Now, by considering Eqs.~\ref{ell0} and \ref{gammas}, one may write the full temperature dependence of $\gamma_s\equiv\gamma_s(T)$ as
\begin{equation}
\gamma_s = \dfrac{(\ell_s-\ell_b) e^{\Delta E/k_BT}}{\ell_s+\ell_be^{\Delta E/k_BT}}~~,
\label{gammas2}
\end{equation}
so that the explicit temperature-dependent expression for the elastic modulus given by Eq.~\ref{elasticmodulus-total} can be written as
\begin{equation}
G(T)=\dfrac{k_BT}{A\Delta\ell^2}\left(\ell_s+\ell_be^{\Delta E/k_BT}\right)\left(1+e^{-\Delta E/k_BT}\right)~~.
\label{elasticmodulus-total2}
\end{equation}

\begin{figure}[!t]
	\centering
	\resizebox{0.45\textwidth}{!}{%
		\includegraphics{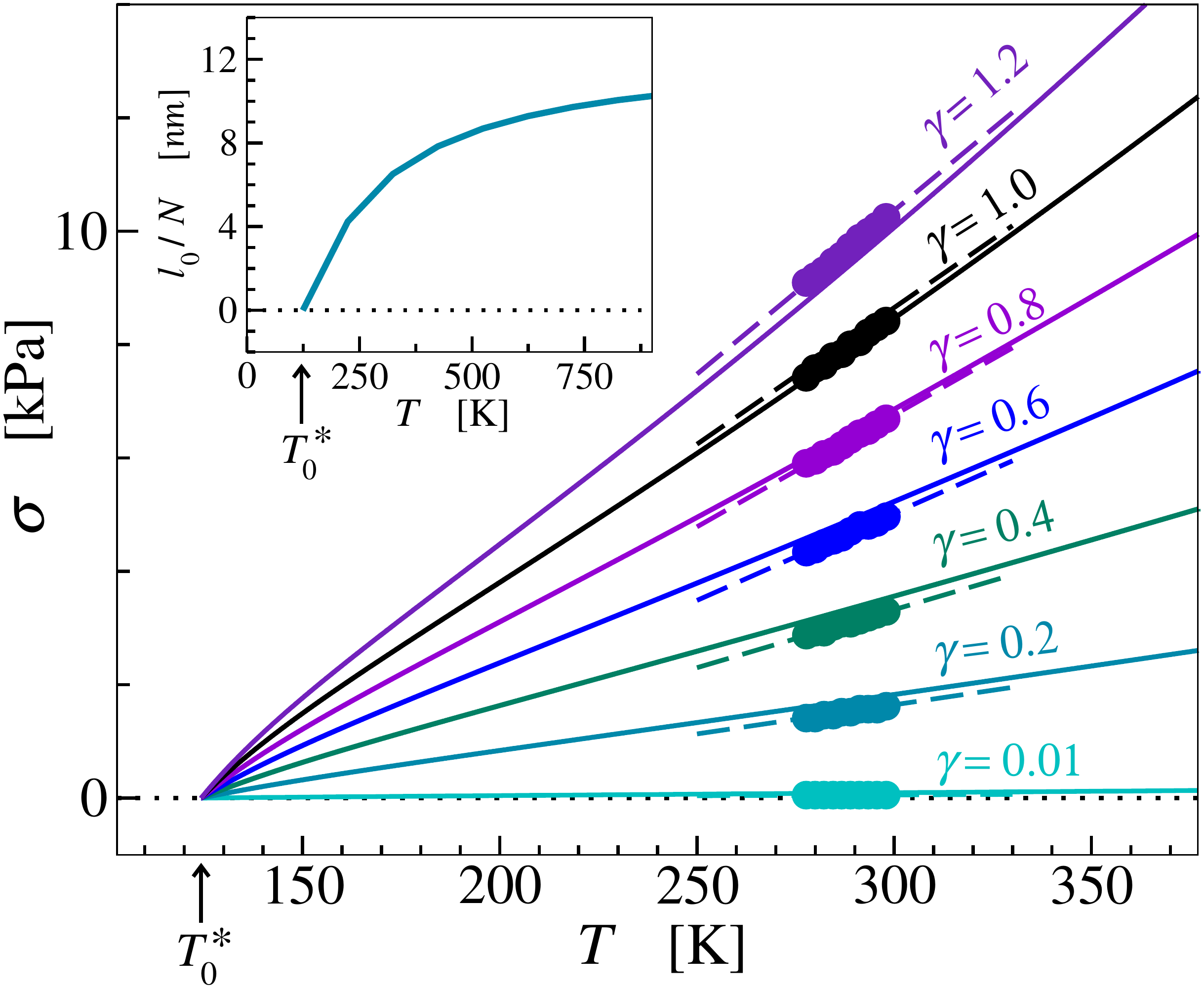}
	}
	\caption{
	Stress $\sigma(T,\gamma)$ as a function of the temperature $T$ for a rubber-like gel subjected to different strains $\gamma$.
	Filled symbols correspond to the experimental data extracted from Ref.~\cite{yoshikawa2021prx}.
	Continuous lines correspond to Eq.~\ref{sigma-gamma} evaluated at the different values of strain $\gamma$, while dashed lines denote the linear phenomenological expression of Ref.~\cite{yoshikawa2021prx}, {\it i.e.}, $G(T)$ given by Eq.~\ref{heurG} so that $\sigma(T,\gamma)= a\,(T-T_0) \gamma$, with similar values of $a= (5.0 \pm 0.2)\times 10^{-2}\,$kPa/K and $T_0= (131\pm8)\,$K.
	Inset panel displays the reduced natural end-to-end distance $\ell_0 / N$ as a function of the temperature $T$. 
	All the theoretical results evaluated with Eq.~\ref{sigma-gamma} presented here were obtained with $\Delta E=3.4\,$pN.nm, $\ell_s=29\,$nm, $\ell_b=-4\,$nm ({\it i.e.}, $\Delta \ell=33\,$nm), and $A=11.575\,$nm$^2$, which yields the temperature $T_0^* = 124.3\,$K where $\ell^*_0\equiv \ell(N,T^*_0,0) = 0$ and $\sigma(T_0^*,\gamma) = 0$. 
	}
	\label{fig:stress-temp}       
\end{figure}

\subsection{Stress $\sigma(T,\gamma)$ vs. temperature}

	In order to validate our approach, we include in Fig.~\ref{fig:stress-temp} a comparison between the stress-strain relation given by Eq.~\ref{sigma-gamma} and the experimental data extracted from Ref.~\cite{yoshikawa2021prx} for tetra-PEG hydrogels.
	By trying several sets of parameters, we found that the parameters $\Delta E = 3.4\,$pN.nm, $\ell_s = 29\,$nm, $\ell_b = -4\,$nm, and $A = 11.575\,\textrm{nm}^2$ are the best ones in order to describe, simultaneously, the whole set of experimental data. 
	Indeed, as it can be seen from the results presented in Fig.~\ref{fig:stress-temp} for $\sigma(T,\gamma)$ as a function of the temperature $T$ and several different values for the strain $\gamma$, the stress-strain relation given by Eq.~\ref{sigma-gamma} obtained here displays a good agreement when compared to the data extracted from the experiments.

	Also, the results presented in Fig.~\ref{fig:stress-temp} indicates that $\sigma(T,\gamma)$ may not display a linear behaviour as a function of the temperature $T$, as proposed in Ref.~\cite{yoshikawa2021prx} and expected from the phenomenological model defined by Eq.~\ref{heurG}.
	In fact, Eq.~\ref{sigma-gamma} indicates that this dependence is slightly non-linear since the parameters $\gamma_s$ and $\gamma_b$ inside the logarithm term are defined by Eqs.~\ref{gammab} and \ref{gammas2}, with both displaying a non-linear dependence on the temperature.

\subsection{Temperature $T_0^*$}
\label{Tzero}

	Interestingly, Fig.~\ref{fig:stress-temp} indicates that the stress-strain relation given by Eq.~\ref{sigma-gamma} displays a characteristic temperature $T_0^{*}$ where $\sigma(T_0^{*},\gamma)=0$.
	Although it might be unphysical to reach that temperature in practice because many solvents like water will be frozen at such a low temperature, we discuss the condition where the temperature $T$ is equal to $T_0^{*}$ since it might shed light on the phenomenological temperature $T_0$ defined in Eq.~\ref{heurG} (see also Ref.~\cite{yoshikawa2021prx}).

	We first note that having $\sigma(T_0^{*},\gamma)=0$ is the same as having that the force $f$ equal to zero at this specific temperature, since the force and stress are related as $f=A\sigma$.
	Then, as suggested in the inset of Fig.~\ref{fig:stress-temp}, where $\ell_0 / N$ is plotted as a function of the temperature $T$, we argue that the condition $f \approx 0$ ({\it i.e.}, $\sigma \approx 0$) will be satisfied when the natural end-to-end distance tends to zero at the temperature $T_0^*$, that is, $\ell_0^*\equiv \ell(N,T^*_0,0) \approx 0$.
	In particular, by imposing that $\ell_0=0$ and considering Eq.~\ref{ell0}, one finds the temperature
\begin{equation}
T_{0}^{*} =  \frac{\Delta E }{k_B  \ln \left(- \ell_s/ \ell_b \right) }~~,
\label{T0temp}
\end{equation} 
which will be positive if $\Delta E>0$, $\ell_b < 0$, and $\ell_s > |\ell_b|$, as in the case we are considering here. 
	At this temperature, one has that the ratio determined by Eq.~\ref{gammab} is given by $\gamma_s(T_{0}^{*})/\gamma_b(T_{0}^{*}) = -\ell_s/\ell_b$.
	Also, from Eq.~\ref{gamma-strain}, one has that $\ell^*\equiv \ell(N,T^*_0,f) = (1+\gamma) \ell^*_0 \approx 0$.
	Hence, the force computed by Eq.~\ref{force-dist} should be given by $f(T_{0}^{*},\ell^{*}) = (\Delta E/\Delta \ell) - (k_BT_{0}^{*} / \Delta \ell) \ln \left( - \ell_s/ \ell_b \right)$, which is zero according to the $T_{0}^{*}$ defined by expression~\ref{T0temp}.
	This means that, at the temperature $T^*_0$, the stress $\sigma(T_{0}^{*},\gamma)$ must be zero as well. 
	In fact, the condition where $\sigma(T_0^{*},\gamma) = 0$ can be possible if the stress in Eq.~\ref{sigma-gamma} is independent of strain. 
	It can happen, in particular, if the argument inside the logarithm term in expression~\ref{sigma-gamma} tends to a finite constant value while $\gamma_s(T_0^*) \gg \gamma$ and $\gamma_b(T_0^*) \gg \gamma$.
	In that case, one will have $(\gamma_b(T_0^*)+\gamma)/(\gamma_s(T_0^*)-\gamma) \approx \gamma_b(T_{0}^{*})/\gamma_s(T_{0}^{*}) \approx -\ell_b/\ell_s$, where in the last step we make use of Eq.~\ref{gammab}. 
	As one may check, by replacing this result and $T_0^{*}$ given by Eq.~\ref{T0temp} in Eq.~\ref{sigma-gamma}, it yields the behaviour observed in Fig.~\ref{fig:stress-temp}, that is, $\sigma(T_0^{*},\gamma) = 0$ at $T_0^{*}=124.3\,$K.

\subsection{$S$ and $E$ contributions}
\label{S-and-E-contrib}

	As in Ref.~\cite{yoshikawa2021prx}, we assume that 
both the stress and the elastic modulus can be defined
as a sum of two contributions, that is,
\begin{equation}
\sigma(T,\gamma) = \sigma_S(T,\gamma) + \sigma_E(T,\gamma)~~,
\label{sigma-total}
\end{equation}
and 
\begin{equation}
G(T) = G_S(T) + G_E(T)~~.
\label{G-total-def}
\end{equation}
	In principle, the authors of Ref.~\cite{yoshikawa2021prx} used the subscripts $S$ and $E$ to denote entropic and energetic contributions, respectively. Although we keep that notation to facilitate the comparison to experiments, their approach indicate that, 
	in contrast to the conventional definitions used in the literature~\cite{flory1953book},
both $\sigma_E(T,\gamma)$ and $G_E(T)$ should include not only terms which are related to the derivative of the internal energy but also terms related to the derivative of the 
	entropy, that is, $f_{E}=A \sigma_E \neq (\partial U/\partial \ell)_{N,T}$ (see Appendix~\ref{definitions-fe-fs} for more details).
	Below we explicitly derive the expressions for both contributions.

\subsubsection{Elastic moduli}

	The $S$ contribution to the elastic modulus can be evaluated from Eq.~\ref{elasticmodulus-total2} using the relation~\cite{yoshikawa2021prx}
\begin{equation}
G_S(T) = T \, \frac{d G(T)}{d T}~~,
\end{equation}
which yields
\begin{eqnarray}
G_S(T) &=& \dfrac{k_BT}{A\Delta\ell^2}\left(\ell_s+\ell_be^{\Delta E/k_BT}\right)\left(1+e^{-\Delta E/k_BT}\right)\nonumber\\
&-& \dfrac{\Delta E}{A\Delta \ell^2}\left(\ell_be^{\Delta E/k_BT} - \ell_se^{-\Delta E/k_BT}\right)~~.
\label{GSofT}
\end{eqnarray}
	The first term in Eq.~\ref{GSofT} is just the elastic modulus $G(T)$ given by Eq.~\ref{elasticmodulus-total2}, thus, by considering Eq.~\ref{G-total-def}, the complementary contribution is easily obtained as
\begin{eqnarray}
G_E(T) = \dfrac{\Delta E}{A\Delta \ell^2}\left(\ell_be^{\Delta E/k_BT} - \ell_se^{-\Delta E/k_BT}\right)~~.
\label{GEofT3}
\end{eqnarray}

	Clearly, if $\Delta E=0$, there will be no energy-related contribution to the elastic modulus from $G_E(T)$, however, if one assumes that $\Delta E>0$, $\ell_s > 0$, and $\ell_b < 0$, Eq.~\ref{GEofT3} indicates that $G_E(T)$ is a negative quantity for all values of $T > T^*_0$. 
	In addition, by considering relatively small strains, linearized response functions can be obtained for the $S$ and $E$ contributions to the stress defined by Eq.~\ref{sigma-total}, that is,
\begin{equation}
\sigma_S(T,\gamma)\simeq G_S(T)\gamma~~,
\label{SigS}
\end{equation}
and
\begin{equation}
\sigma_E(T,\gamma)\simeq G_E(T)\gamma~~.
\label{SigE}
\end{equation} 
	Hence, if $G_E(T)<0$, one will observe a negative contribution to the stress, {\it i.e.}, $\sigma_E(T,\gamma)<0$, while the $S$ contribution to the stress, $\sigma_S(T,\gamma)$, will be positive for all $T> T^*_0$, since $\sigma(T,\gamma) \simeq G(T)\gamma$ must be also positive.
	In order to verify this behaviour for the general case, that is, not only at the linear response regime but also at higher values of $\gamma$, next we calculate the full expressions of $\sigma_S(T,\gamma)$ and $\sigma_E(T,\gamma)$.

\subsubsection{Positive ($S$) contribution to the stress $\sigma(T,\gamma)$}

	Following the approach presented in Ref.~\cite{yoshikawa2021prx}, the $S$ contribution to the stress can be evaluated as
\begin{equation}
\sigma_S(T,\gamma) = T \left[ \frac{\partial }{\partial T}\sigma(T,\gamma)\right]_{\gamma}~~,
\end{equation}
so that Eq.~\ref{sigma-gamma} yields
\begin{eqnarray}
\sigma_S(T,\gamma)&=&\dfrac{k_BT}{A\Delta\ell}\ln\left(\dfrac{\gamma_b+\gamma}{\gamma_s - \gamma}\right)\nonumber\\
&+& \alpha(T)\frac{ k_BT^2}{A\Delta \ell}\left(\frac{\gamma_s+1}{\gamma_s-\gamma} - \frac{\gamma_b-1}{\gamma_b + \gamma}\right)~~,
\label{sigmaS}
\end{eqnarray}
with $\gamma_b$ and $\gamma_s$ given by Eqs.~\ref{gammab} and~\ref{gammas2}, respectively, whereas
\begin{equation}
\alpha(T) = \frac{1}{\ell_0} \left( \frac{d \ell_0}{d T} \right)_{N} =  \frac{\Delta E}{A\Delta\ell}\dfrac{1}{TG(T)}~~
\label{alphaTE}
\end{equation}
is the coefficient of thermal expansion of the chains in the gel, with $G(T)$ being the elastic modulus given by Eq.~\ref{elasticmodulus-total2}.

\begin{figure}[!t]
\centering
\resizebox{0.45\textwidth}{!}{%
\includegraphics{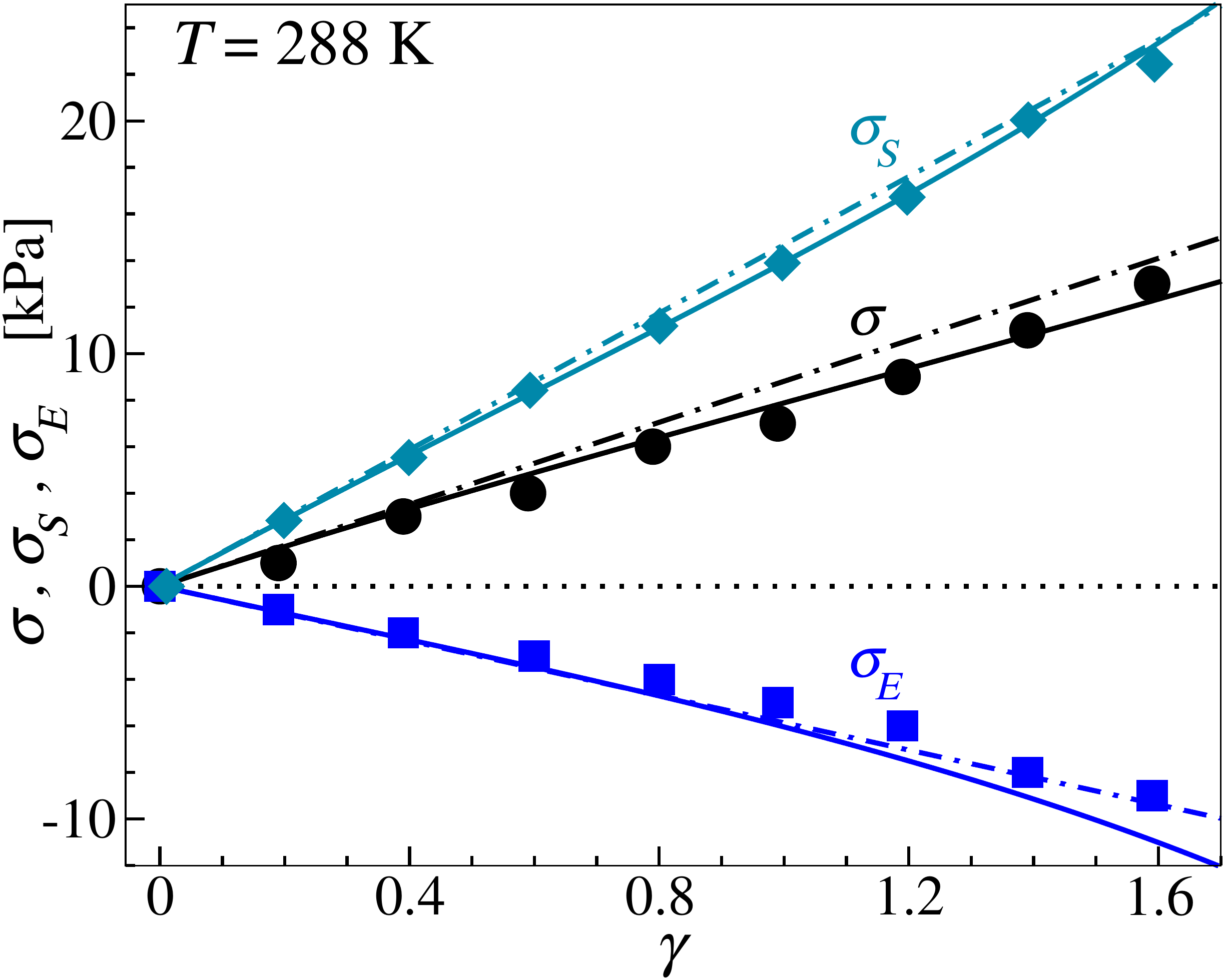}
}
\caption{Filled symbols indicate the experimental data extracted from Ref.~\cite{yoshikawa2021prx} and correspond to the stresses as functions of the strain evaluated at the temperature $T=288\,$K.
	The black lines denote the total stress $\sigma(T,\gamma)$, while the teal and the blue lines correspond to its positive, $\sigma_S(T,\gamma)$, and negative, $\sigma_E(T,\gamma)$, contributions, respectively.
	Continuous lines denote the non-linear expressions given by Eqs.~\ref{sigma-gamma},~\ref{sigmaS}, and~\ref{sigmaE}, while the dash-dotted lines denote the linearized relations given by Eqs.~\ref{sigma-gamma-linear},~\ref{SigS}, and~\ref{SigE}, with the corresponding elastic moduli evaluated through Eqs.~\ref{elasticmodulus-total2},~\ref{GSofT}, and~\ref{GEofT3}.
	Here we consider the same parameters used in Fig.~\ref{fig:stress-temp}, that is, $\Delta E=3.4\,$pN.nm, $\ell_s=29\,$nm, $\ell_b=-4\,$nm ({\it i.e.}, $\Delta \ell=33\,$nm), and $A=11.575\,$nm$^2$.
}
\label{fig:stress-gamma}  
\end{figure}

	It is worth mentioning that, under tension, {\it i.e.,} for $f\ne 0$, the thermal expansion behaviour of the chains is governed by $\alpha(T,f) = \ell^{-1} [\partial \ell/ \partial T]_{N,f}$,
with the end-to-end distance $\ell$ given by Eq.~\ref{end-to-end}, 
which can be explicitly evaluated as
\begin{equation}
\alpha(T,f) = \frac{(\Delta E - f\Delta\ell)\Delta\ell}{k_BT^2} 
\frac{ [1+e^{-(\Delta E - f\Delta\ell)/k_BT}]^{-1} }{ [\ell_s+\ell_be^{(\Delta E - f\Delta\ell)/k_BT}] ~~} \,.
\end{equation}
	Thus, if $f >\Delta E/\Delta\ell$, it is possible that the chains react as predicted by the Gough-Joule effect with $\alpha(T,f)$ being a negative function, {\it i.e.}, it is possible that tensioned chains contract when the temperature is increased, as it seems to occur in tetra-PEG hydrogels explored in Ref.~\cite{yoshikawa2021prx}.

\subsubsection{Negative ($E$) contribution to $\sigma(T,\gamma)$}

	Now, by considering Eq.~\ref{sigma-total}, that is, $\sigma_E(T,\gamma) = \sigma(T,\gamma) - \sigma_S(T,\gamma)$, one can consider Eqs.~\ref{sigma-gamma} and~\ref{sigmaS} to identify the negative contribution to the stress as
\begin{equation}
\sigma_E(T,\gamma) = \frac{\Delta E}{A\Delta \ell} -\alpha(T)\frac{ k_BT^2}{A\Delta \ell}\left(\frac{\gamma_s + 1}{\gamma_s-\gamma} - \frac{\gamma_b-1}{ \gamma_b + \gamma}\right)~,
\label{sigmaE}
\end{equation}
with the usual definitions of $\gamma_b$ (Eq.~\ref{gammab}) and $\gamma_s$ (Eq.~\ref{gammas2}), and $\alpha(T)$ given by Eq.~\ref{alphaTE}.

\subsection{Stresses $\sigma$, $\sigma_S$, and $\sigma_E$ vs. strain}
	
	Figure~\ref{fig:stress-gamma} includes the comparison between our theoretical results and the experimental data extracted from Ref.~\cite{yoshikawa2021prx} determined for tetra-PEG hydrogels at a constant temperature.
	In Fig.~\ref{fig:stress-gamma} one sees how well the non-linear relations found for the stresses, given by Eqs.~\ref{sigma-gamma}, \ref{sigmaS} and~\ref{sigmaE}, describe the experimental results obtained as a function of the strain $\gamma$ at $T=288\,$K.
	This corroborates expression~\ref{sigma-gamma} for the total stress and validates its consequences, \textit{i.e.}, Eqs.~\ref{sigmaS} and~\ref{sigmaE}, which were determined for its $S$ and $E$ contributions, respectively.
 	Importantly, the results in Fig.~\ref{fig:stress-gamma} show that, even when the full dependence of the contribution $\sigma_E(T,\gamma)$ to the stress as a function of $\gamma$ is considered, a negative elastic contribution is observed for strains $\gamma$ as higher as 1.6, where the system is already at the non-linear response regime.

	Furthermore, the results presented in Fig.~\ref{fig:stress-gamma} indicate that, when the positive ($S$) and the negative ($E$) contributions to the elastic modulus are given by Eqs.~\ref{GSofT} and~\ref{GEofT3}, respectively, the linearized relations (dash-dotted lines) evaluated through Eqs.~\ref{SigS} and~\ref{SigE} also describe reasonably well the experimental data.
	Even so, it is worth noting that Eqs.~\ref{sigma-gamma-linear}, \ref{SigS}, and \ref{SigE} are valid only if $\gamma < \gamma_s$, since the non-linear expressions for 
 $\sigma(T,\gamma)$, $\sigma_S(T,\gamma)$, and $\sigma_E(T,\gamma)$, all diverge for $\gamma = \gamma_s$. 
	Thus, the value of $\gamma_s$ establishes the maximum strain that the chains in the network can be submitted at a temperature $T$.

\subsection{Stresses $\sigma$, $\sigma_S$, and $\sigma_E$ vs. temperature}

\begin{figure}[!t]
\centering
\resizebox{0.45\textwidth}{!}{%
\includegraphics{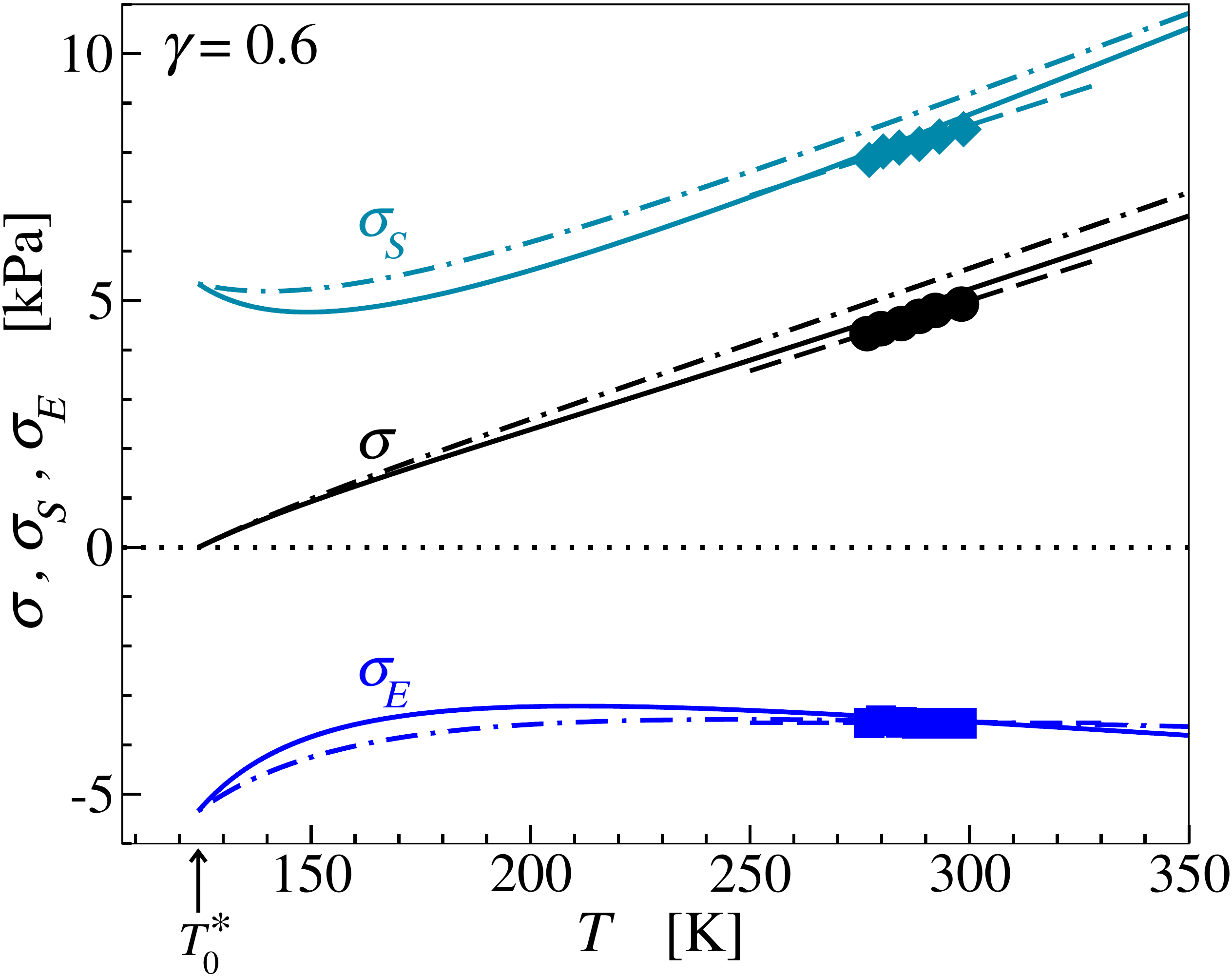}
}
\caption{Filled symbols indicate the experimental data extracted from Ref.~\cite{yoshikawa2021prx} and correspond 
to the stresses as functions of the temperature $T$ evaluated at constant strain, $\gamma=0.6$.
	Black circles correspond to the total stress $\sigma(T,\gamma)$, while the teal diamonds and the blue squares denote, respectively, the positive, $\sigma_S(T,\gamma)$, and negative, $\sigma_E(T,\gamma)$, contributions.
	Continuous lines denote the non-linear expressions obtained from our model, that is, Eqs.~\ref{sigma-gamma},~\ref{sigmaS}, and~\ref{sigmaE}, while the dash-dotted lines correspond to the linearized relations, {\it i.e.}, Eqs.~\ref{sigma-gamma-linear},~\ref{SigS}, and~\ref{SigE}, with the respective elastic moduli given by Eqs.~\ref{elasticmodulus-total2},~\ref{GSofT}, and~\ref{GEofT3}, all plotted with the same parameters used in Figs.~\ref{fig:stress-temp} and~\ref{fig:stress-gamma}, that is, $\Delta E=3.4\,$pN.nm, $\ell_s=29\,$nm, $\ell_b=-4\,$nm, ({\it i.e.}, $\Delta \ell=33\,$nm), and $A=11.575\,$nm$^2$, which leads to $T_0^* = 124.3\,$K (Eq.~\ref{T0temp}).
	Straight dashed lines correspond to the phenomenological linear behaviour described by $\sigma(T,\gamma)= a\, (T-T_0) \gamma$ proposed in Ref.~\cite{yoshikawa2021prx}, where $G(T)$ is given by Eq.~\ref{heurG}, from where one gets $\sigma_S(T,\gamma) = a\,T \gamma$ and $\sigma_E(T,\gamma) = -a\,T_0 \gamma$, with parameters $a = 4.75\times 10^{-2}\,$kPa/K and $T_0 = 124.6\,$K.
}
\label{fig:stresses-temp}
\end{figure}

	Now, in order to illustrate how the stress and its positive ($S$) and the negative ($E$) contributions behave as functions of the temperature but at a fixed strain, we include in Fig.~\ref{fig:stresses-temp} a comparison between our theoretical results and the experimental data obtained for tetra-PEG hydrogels extracted from Ref.~\cite{yoshikawa2021prx}.
	Accordingly, one can observe a good agreement between the experimental data (filled symbols) and our non-linear expressions (continuous lines), \textit{i.e.}, Eqs.~\ref{sigma-gamma}, \ref{sigmaS} and \ref{sigmaE}.
	Although qualitatively, a similar behaviour is observed for the linearized relations (dash-dotted lines), that is, Eqs.~\ref{sigma-gamma-linear}, \ref{SigS}, and \ref{SigE}.
	Besides the results obtained from our model (continuous and dash-dotted lines), Fig.~\ref{fig:stresses-temp} also includes the results obtained with the phenomenological model (dashed straight lines) proposed in Ref.~\cite{yoshikawa2021prx}, where $\sigma(T,\gamma)= G(T) \gamma$, with $G(T)$ defined by Eq.~\ref{heurG}.

	We note that, in contrast to what is expected from the phenomenological model, the positive contribution to the stress, {\it i.e.}, $\sigma_S(T,\gamma)$ given by Eq.~\ref{SigS}, is not a linear function of the temperature.
	In that model, the $S$ contribution to the stress is a bi-linear function of the temperature and the strain, that is, $\sigma_S(T,\gamma) = G_S(T)\gamma = aT\gamma$.
	As a consequence, the $S$ contribution to the elastic modulus is simply $G_S(T) = aT$, which is clearly different from what is found from our model, where both $\sigma_S(T,\gamma)$ and $G_S(T)$ are more complicated functions of the temperature.
	Furthermore, as shown in Fig.~\ref{fig:stresses-temp}, the negative contribution to the stress, $\sigma_E(T,\gamma)$, and to the elastic modulus, $G_E(T)$, in our model are also temperature-dependent, while in the phenomenological model those contributions are constant through temperature variations, \textit{i.e.}, $\sigma_E(T,\gamma) = G_E(T)\gamma = - aT_0\gamma$.

	Yet, one might expect that, even with such differences, for  small strains and relatively high temperatures, both models might describe the experimental data well.
	As we discuss next, such equivalence can be used to establish the dependence of the parameters defined in our model and the experimental conditions.

\subsection{Relationship between our model and the 
phenomenological model of Ref.~\cite{yoshikawa2021prx}}
	
	To understand how the parameters of our model might be related to the model and results presented in Refs.~\cite{yoshikawa2021prx}, we first consider Eq.~\ref{sigma-gamma} for small strains and relatively high temperatures,  \textit{i.e.}, the case where $\gamma/\gamma_s \ll 1$ and $\Delta E/k_BT$ $\ll 1$. 
	In this case, one can assume Eq.~\ref{sigma-gamma-linear} to be valid, so that the elastic modulus, given by Eq.~\ref{elasticmodulus-total2}, can be linearized by assuming that $e^x \simeq 1 + x$. 
	Hence, by neglecting second-order terms in $x=\Delta E/k_BT$, one finds that
\begin{equation}
G(T) \simeq \dfrac{\ell_s+\ell_b}{\ell_s-\ell_b}\dfrac{2k_B}{A(\ell_s-\ell_b)}\left[T - \dfrac{\ell_s-\ell_b}{\ell_s+\ell_b}\dfrac{\Delta E}{2k_B}\right]~~.
\label{sigma-gamma-assynt}
\end{equation} 
	Remarkably, with such linearization, the approximated elastic modulus $G(T)$ obtained from our model seems to exhibit the same linear behaviour of the phenomenological model that was proposed in Ref.~\cite{yoshikawa2021prx}. 
	Thus, by comparing Eq.~\ref{heurG} and Eq.~\ref{sigma-gamma-assynt}, one finds the following relations
\begin{equation}
a = \dfrac{\ell_s+\ell_b}{\ell_s-\ell_b}\dfrac{2k_B}{A(\ell_s-\ell_b)}
\label{a-A-ls-lb}
\end{equation}
and
\begin{equation}
T_0 =\dfrac{\ell_s-\ell_b}{\ell_s+\ell_b}\dfrac{\Delta E}{2k_B}~~.
\label{T0-dE-ls-lb}
\end{equation}

	Now, it is instructive to inspect what happens if the parameters obtained from the heuristic non-linear fit of Eqs.~\ref{sigma-gamma},~\ref{sigmaS}, and~\ref{sigmaE} to the data presented in Figs.~\ref{fig:stress-temp},~\ref{fig:stress-gamma}, and~\ref{fig:stresses-temp} are plugged into the above expressions.
	In particular, by assuming $\Delta E=3.4\,$pN.nm, $\ell_s=29\,$nm, $\ell_b=-4\,$nm, and $A=11.575\,$nm$^2$, one finds from the above relations that $a = 5.47\times10^{-2}\,$kPa/K and $T_0 = 162.6\,$K, which overestimates the values $a = (5.0 \pm 0.2)\times10^{-2}$ kPa/K and $T_0 = (131 \pm 8)\,$K obtained from the fit of the data to the linear expressions of the  phenomenological model~\cite{yoshikawa2021prx}, {\it i.e.,}~those determined from Eq.~\ref{heurG}.
	Nevertheless, because this correspondence is not valid in general, but only in a limiting case, one may not expect that these relations will give consistent values for $a$ and $T_0$. 
	Indeed, the temperature $T_0 = 162.6\,$K (Eq.~\ref{T0-dE-ls-lb}) estimated from the parameters obtained with the non-linear expressions is different not only from the characteristic temperature $T_0^* = 124.3\,$K (Eq.~\ref{T0temp}) of our non-linear model, but also from the value determined  from the phenomenological model, {\it i.e.}, $T_0 = (131 \pm 8)\,$K.
	Interestingly, the stresses obtained through the linear approximation of $G(T)$ given by Eq.~\ref{sigma-gamma-assynt} seems to fit the experimental data much better if small changes are made to the parameters.
	In particular, by considering $\Delta E = 2.7\,$pN.nm, $\ell_s = 30\,$nm, $\ell_b = -4.35$ nm (\textit{i.e.}, $\Delta\ell = 34.35\,$nm), and $A = 11.575\,\textrm{nm}^2$, the relations~\ref{a-A-ls-lb} and~\ref{T0-dE-ls-lb} yield  $a = 5.18\times10^{-2}\,$kPa/K and $T_0 = 131\,$K, which are closer to the parameters obtained in Ref.~\cite{yoshikawa2021prx} and that were used to obtain the curves for different strains presented in Fig.~\ref{fig:stress-temp} as the dashed lines.
	However, it is important noting that, if this last set of parameters is plugged into the non-linear expression, Eq.~\ref{sigma-gamma}, one gets $T_0^*=101.32\,$K (Eq.~\ref{T0temp}) and the agreement with experimental data is poor, which indicates that the equivalence between the $G(T)$ given by Eq.~\ref{sigma-gamma-assynt} and the elastic modulus of the phenomenological model (Eq.~\ref{heurG}) is somewhat fortuitous due to the linearization procedure.

	As it can be inferred from Refs.~\cite{yoshikawa2021prx,sakumichi2021polymj}, the experimental results presented in Figs.~\ref{fig:stress-temp},~\ref{fig:stress-gamma}, and~\ref{fig:stresses-temp} were all measured at the same experimental condition, which is characterized by gels with network connectivity $p = 0.915$ assembled from tetra-functional precursor molecules with molar mass $M = 20\,$kg/mol at a concentration $c=60\,$kg/m$^3$.
	In the following we will explore an analysis based on the experimental data extracted from Refs.~\cite{yoshikawa2021prx,sakumichi2021polymj} in order to understand how the parameters $\ell_s$, $\ell_b$, $\Delta E$, and $A$ might depend on $p$, $M$, and $c$.

\begin{figure*}[!t]
	\centering
	\resizebox{0.97\textwidth}{!}{%
		\includegraphics{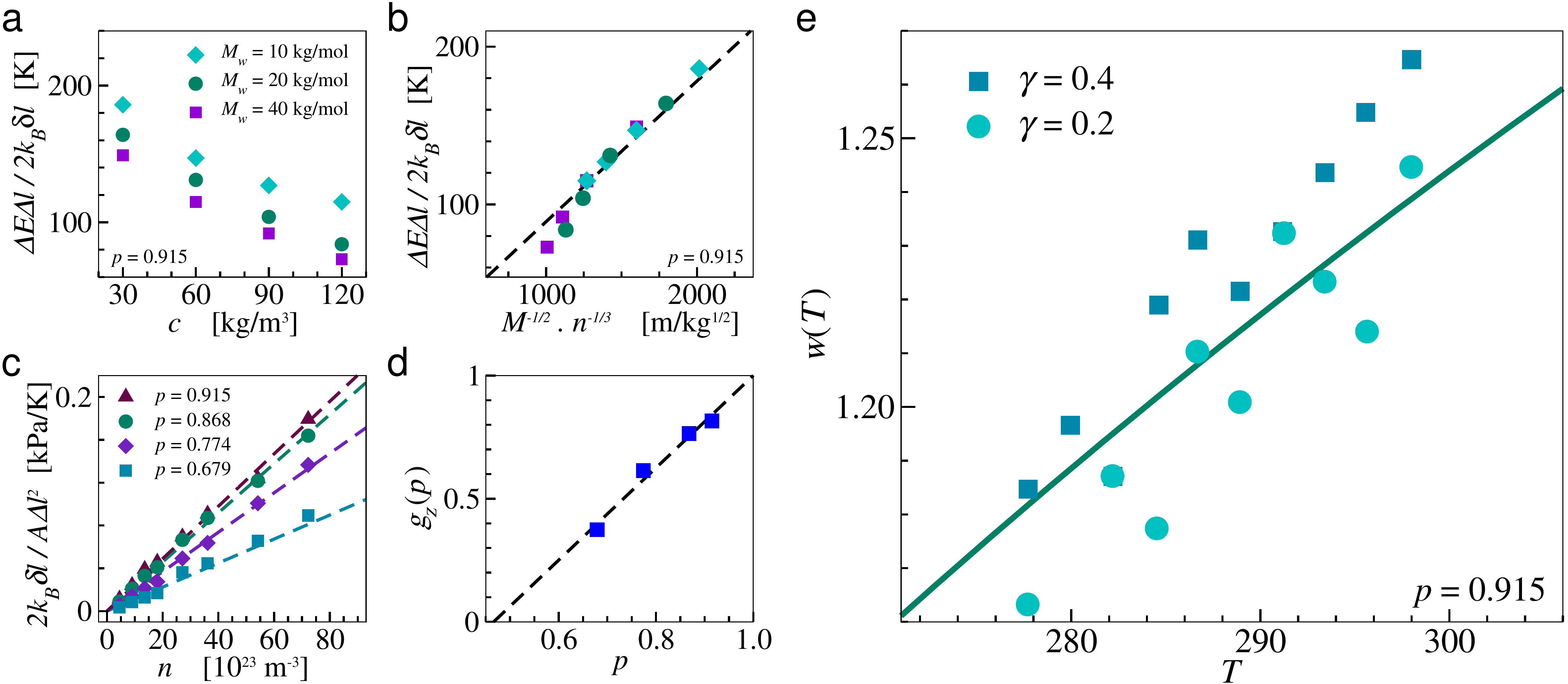}
	}
	\caption{Analysis from the experimental data extracted from Ref.~\cite{yoshikawa2021prx} for tetra-PEG ($z=4$) hydrogels.
	(a)~Temperature $T_0 = \Delta E\Delta\ell / 2k_B\delta\ell$, Eq.~\ref{T0-dE-ls-lb}, as a function of the concentration $c$ of precursor molecules with different molar masses $M$.
	(b)~Master curve of the data presented in (a) showing that $T_0=\Delta E\Delta\ell / 2k_B\delta\ell\propto M^{-1/2}n^{-1/3}$. 
	(c)~Parameter $a = 2k_B\delta \ell/A\Delta\ell^2$, Eq.~\ref{a-A-ls-lb}, as a function of the number density of precursor molecules $n=cN_A/M$, Eq.~\ref{numberdensity}, for different values of the connectivity $p$.
	Dashed lines corresponds to linear functions obtained from the fit of the expression $a = w^*g_z(p)k_Bn$, Eq.~\ref{a-wstar-gzp-n}, to the experimental data obtained for each value of $p$.
	(d)~Filled symbols correspond to the values of the function $g_z(p)$ found from the linear fit of Eq.~\ref{a-wstar-gzp-n} to the data presented in (c), and assuming that $w^* = 2.17$.
	The dashed line corresponds to the fit of the linear function $g_z(p)$ given by Eq.~\ref{gz} imposing that $g_z(1)=1$, which yields $p_z^{*}=0.47$.
	Panel (e) includes results for the temperature-dependent correction factor $w(T) = G(T)/g_z(p)nk_BT$.
	Filled symbols correspond to the experimental data of Fig.~\ref{fig:stress-temp}, with the values of the elastic modulus estimated as $G(T)\approx \sigma(T,\gamma)/\gamma$ for both $\gamma = 0.2$ and $\gamma = 0.4$, and considering the value of $g_z(p)$ given by Eq.~\ref{gz} with $p_z^*\simeq 0.47$ for the connectivity $p = 0.915$,  molar mass $M = 20~\textrm{kg/mol}$, and concentration $c = 60~\textrm{kg/m}^3$, while the continuous line indicates the theoretical estimate given by Eq.~\ref{wT2} with $\Delta E=3.4\,$pN.nm, $\ell_s=29\,$nm, $\ell_b=-4\,$nm, ({\it i.e.}, $\Delta \ell=33\,$nm),  $A=11.575\,$nm$^2$, and $w^*=2.17$.
}
	\label{fig:temp0Xconc_temp0Xmol} 
\end{figure*}

\subsubsection{Energy difference $\Delta E$ and temperature $T_{0}^{*}$}

	We first consider how the energy difference $\Delta E$ changes for different values of $p$, $M$, and $c$.
	In fact, in the following we replace the concentration of precursor molecules $c$ by its number density using the relation
\begin{equation}
n = c\, \frac{N_A}{M} ~~,
\label{numberdensity}
\end{equation}
where $N_A$ is the Avogadro's constant.
	It was established from experimental data analyses in Refs.~\cite{yoshikawa2021prx,sakumichi2021polymj} that $T_0\equiv T_0(c,M)$, \textit{i.e.}, the phenomenological temperature seems to be independent of the connectivity $p$.
	Also, it was suggested in Ref.~\cite{yoshikawa2021prx} that $T_0\propto M^{-1/2}\,n^{-1/3}$ for lower values of $M$ and $n$.
	Now, since Eq.~\ref{T0-dE-ls-lb} establishes that $\Delta E\propto T_0$, one needs to discuss how $\delta\ell / \Delta\ell$ depends on the experimental conditions, with $\delta\ell = \ell_s+\ell_b$ and $\Delta\ell = \ell_s-\ell_b$.
	In particular, by assuming that both $\ell_s$ and $\ell_b$ display the same dependency\footnote{The specific dependence of $\ell_s$ and $\ell_b$ on $M$ might be inferred from known polymer models~\cite{rubinsteinbook} such as, {\it e.g.}, the Gaussian chain, the worm-like chain (WLC), the and freely-jointed chain (FJC).
	For instance, one can assume $M_f = 2(M/z)$ as the molar mass of bridged chains between crosslinks of functionality $z$, so that the lengths $\ell_s$ and $\ell_b$ would follow the end-to-end distance dependence on $M_f$, which is a power-law of the type $\ell\propto (M_f)^\nu$, with $\nu$ being the Flory exponent~\cite{flory1953book}.}
	on the molar mass $M$ and no dependency on the number density of precursor molecules $n$, one finds that the ratio $\delta\ell/\Delta\ell$ will be independent of $M$ and $n$.
	Hence, one can infer that the energy difference behaves in the same manner as $T_0$, that is, $\Delta E \propto M^{-1/2}n^{-1/3}$.
	In Fig.~\ref{fig:temp0Xconc_temp0Xmol}(a) we include the right side of Eq.~\ref{T0-dE-ls-lb} as a function of the concentration $c$ for different values of $M$, while in
Fig.~\ref{fig:temp0Xconc_temp0Xmol}(b) we indicate that $\Delta E$, just as $T_0$, should be indeed proportional to $M^{-1/2}n^{-1/3}$, as suggested in Ref.~\cite{yoshikawa2021prx}.

	Finally, it is worth noting that, because of the definition of $T_{0}^{*}$, Eq.~\ref{T0temp}, which involves only the ratio $\ell_s/\ell_b$, one will also have $T_{0}^{*} \propto \Delta E$, so that this temperature is expected to display the same dependence on the molar mass and number density of the phenomenological temperature $T_0$, that is, $T_{0}^{*} \propto M^{-1/2}n^{-1/3}$.
	Although systematic procedures based on numerical simulations of polymer systems suggest that, indeed, coarse-grained potentials should depend on the number density~\cite{reith2003jcomputchem}, we are not aware of any quantitative relationship between $\Delta E$ and either $M$ or $n$ that is obtained from first principles.

\subsubsection{Pre-factor of $G(T)$ for different connectivities $p$}

	Accordingly, Ref.~\cite{yoshikawa2021prx} presents an extensive study where experiments with gels formed by networks with tetra-PEG chains were performed in a broad range of experimental conditions.
	In particular, the study considered different connectivities $p$ while keeping the same molar mass $M$ and number density $n$, which allowed us to explore their results in order to infer how the pre-factor related to $A$ in Eq.~\ref{elasticmodulus-total2} depends on $n$, $M$, and $p$.

	From the theoretical point-of-view, diluted networks with different connectivities $p$ can be described by the effective medium theory based on regular networks presented in Ref.~\cite{nishi2015jcp}.
	In this case, the elastic modulus can be written as
\begin{equation}
G(T) = g_z(p) G_c(T) ~~,
\label{GpT}
\end{equation}
with $G_c(T)$ being the elastic modulus of the fully connected network, \textit{i.e.}, $p=1$, and $g_z(p)$ being a function that depends on the connectivity $p$ of the network.
	For sufficiently highly connected networks, the latter can be written as a linear function as~\cite{nishi2015jcp}
\begin{equation}
g_z(p) = \frac{p-p_z^*}{1-p_z^*}~~,
\label{gz}
\end{equation}
where $p^*_z=2/z$ is a parameter related to
the coordination number (or functionality) $z$ of the crosslinks in the idealized complete network ({\it e.g.}, $z=4$ and  $z=6$ for square  and cubic lattices, respectively).
	Since the experiments were performed with the tetra-PEG gels~\cite{sakai2008macromo,yasuda2009macromo} where the precursor molecules present a coordination equal to $z=4$, one expects that $p_z^* = 1/2$ and $g_z(p) =2p-1$. 
	Furthermore, as suggested in Refs.~\cite{yoshikawa2021prx,sakumichi2021polymj}, the elastic modulus of a fully connected gel network can be approximated by
\begin{equation}
G_c(T) = w^* n\,k_B(T-T_0)~~,
\label{Gc}
\end{equation}
where $n$ is the number density given by Eq.~\ref{numberdensity}, $T_0$ is the phenomenological temperature, and $w^*$ is a positive phenomenological parameter that may be found from experimental analysis (in Refs.~\cite{yoshikawa2021prx,sakumichi2021polymj}, the authors found its value to be $2.4$).
	By considering the elastic modulus $G(T)$ given by the phenomenological model, Eq.~\ref{heurG}, and the definitions~\ref{GpT} and~\ref{Gc}, one finds that
\begin{equation}
a = w^*g_z(p) k_B\, n~~,
\label{a-wstar-gzp-n}
\end{equation}
which is corroborated by the linear behaviour observed for $a = 2 k_B \delta \ell / A \Delta \ell^2$ presented in Fig.~\ref{fig:temp0Xconc_temp0Xmol}(c).
	Hence, by taking the data from Ref.~\cite{yoshikawa2021prx} and reanalyzing it consistently with $g_z(1)=1$, we found from Fig.~\ref{fig:temp0Xconc_temp0Xmol}(c) that the phenomenological parameter $w^*$ is closer to $2.17$.
	In Fig.~\ref{fig:temp0Xconc_temp0Xmol}(d) we plot the function $g_z(p)$ found from the fit as a function of $p$, from where one can verify that the linear expression given by Eq.~\ref{gz} describes the experimental data of Ref.~\cite{yoshikawa2021prx} well.
	Also, the fit of Eq.~\ref{gz} to the data presented in Fig.~\ref{fig:temp0Xconc_temp0Xmol}(d) imposing that $g_z(1)=1$ yields $p^*_z \simeq 0.47$, which is remarkably close to the value that is expected for tetra-PEG networks with $z=4$.

	It is worth noting that, through the identity~\ref{a-wstar-gzp-n} and the relation based on the phenomenological model given by Eq.~\ref{a-A-ls-lb}, one finds that
\begin{equation}
A\Delta\ell = \frac{2}{w^*}
 \frac{\delta\ell}{ \Delta\ell }
\frac{1}{g_z(p) n} ~~. 
\label{ADl2}
\end{equation}
	As discussed before, we assume that $\delta \ell/\Delta\ell = (\ell_s+\ell_b)/(\ell_s-\ell_b)$ is independent of $p$ and $M$, thus Eq.~\ref{ADl2} implies that $A\Delta\ell\propto [g_z(p)n]^{-1}$. 
	Even though one cannot tell exactly how the effective area $A$ changes with the different values of $p$, $n$, and $M$, since it is not possible to know exactly how $\Delta\ell$ changes with those quantities, one can infer, at least, that the product of those parameters is inversely proportional to $g_z(p)$ and the number density $n$.

	In addition, one should note that, although we have considered specific values for $\ell_s$, $\ell_b$, and $A$, there are, in fact, only two parameters (besides $\Delta E$) that are relevant to the behaviour of the stresses and the elastic moduli, which are the ratio $q_{\ell}^{\,}=-\ell_s/\ell_b$ and the product $r_{\ell}^{\,}=A \Delta \ell$.
	For instance, by setting $q_{\ell}^{\,}=7.25$ and $r_{\ell}^{\,}=381.975\,$nm$^3$, 
one could have used half of the lengths, that is, $\ell_s=14.5\,$nm and $\ell_b=-2\,$nm, then twice the value of the effective area, {\it i.e.}, $A=23.150\,$nm$^2$, in order to obtain exactly the same curves showed in Figs.~\ref{fig:stress-temp},~\ref{fig:stress-gamma}, and~\ref{fig:stresses-temp}, since the values of $q_{\ell}^{\,}$ and $r_{\ell}^{\,}$ remain the same.
	As it happens, Eq.~\ref{ADl2} establishes an interesting relation between $q_{\ell}^{\,}$ and $r_{\ell}^{\,}$ that can be conveniently explored in further studies to describe other experimental data sets.
	It is worth mentioning that the values $\ell_s=29\,$nm and $\ell_b=-4\,$nm are consistent with
the estimates obtained from PEG chains used in single-molecule force-extension experiments~\cite{kolberg2019JACS}.

\subsection{Temperature-dependent correction factor $w(T)$}

	It is important noting that the authors in Refs.~\cite{yoshikawa2021prx,sakumichi2021polymj} did not explained the origin of the phenomenological correction $w^*$ in Eq.~\ref{Gc}, and so in the expressions~\ref{a-wstar-gzp-n} and~\ref{ADl2}.
	Here we argue that it could be effectively described as a temperature-dependent correction factor to the elastic modulus $G(T)$.
	In particular, we define
\begin{equation}
w(T) \equiv \dfrac{G(T)}{g_z(p)nk_BT}~~,
\label{wT}
\end{equation}
which should be constant if no temperature-dependent correction is needed to describe the experimental data, as in the case of the elastic modulus $G(T)$ obtained considering only the entropic contributions to the system, {\it i.e.}, $G(T) \propto k_B T$.
	As one can see in Fig.~\ref{fig:temp0Xconc_temp0Xmol}(e), we present the plot of the ratio defined by Eq.~\ref{wT} 
as a function of the temperature extracted from the data of Fig.~\ref{fig:stress-temp} (here, only the data that correspond to strains $\gamma = 0.2$ and $\gamma = 0.4$ were considered to calculate the experimental elastic modulus $G(T) \simeq \sigma(T,\gamma)/\gamma$).
	In contrast to what is expected from the purely entropic models, 
Fig.~\ref{fig:temp0Xconc_temp0Xmol}(e) shows that the ratio $w(T)$ is not constant, indicating that a temperature-dependent factor must be present in any modelling approach that aims to describe rubber-like gels.

	We have shown that both the stress-strain relation obtained from our model, Eq. \ref{sigma-gamma}, and its linearized approximation, Eq.~\ref{sigma-gamma-linear}, with the elastic modulus given by Eq.~\ref{sigma-gamma-assynt}, describe the experimental data of Figs.~\ref{fig:stress-temp},~\ref{fig:stress-gamma}, and \ref{fig:stresses-temp}, reasonably well.
	Hence, by assuming that the relationship given by Eq.~\ref{ADl2} is also valid for our model, one has that
\begin{equation}
\frac{1}{A\Delta\ell^2} = \frac{w^*g_z(p) n}{2 \left(\ell_s + \ell_b \right)}~,
\label{overADl}
\end{equation}
which can be replaced in the elastic modulus $G(T)$ given by Eq.~\ref{elasticmodulus-total2}, so that, in principle, the temperature-dependent correction factor can be explicitly written as
\begin{equation}
w(T) = w^* \frac{ \left(1+e^{-\Delta E/k_BT}\right)}{2} \left( \frac{\ell_s + \ell_be^{\Delta E/k_BT}}{\ell_s+\ell_b}\right)
 ~~.
\label{wT2}
\end{equation}
	As one may observe from Fig.~\ref{fig:temp0Xconc_temp0Xmol}(e), the above expression describes the experimental data of Ref.~\cite{yoshikawa2021prx} well.
	For relatively high temperatures, {\it i.e.}, where $x=\Delta E/k_BT \ll 1$ and $e^{x} \simeq 1 + x$, Eq.~\ref{wT2} leads to a temperature-dependent correction factor which is given by
\begin{equation}
w(T) \simeq w^* 
\left( 1 - \frac{\Delta\ell}{2\delta\ell} \frac{\Delta E}{k_BT }
 \right) = w^* \left( 1 - \frac{T_0}{T} \right) ~~,
\end{equation}
where we used the definition of $T_0$ given by Eq.~\ref{T0-dE-ls-lb}.
	Accordingly, by considering the definition of $w(T)$, Eq.~\ref{wT}, the above approximation leads to the phenomenological relation for $G(T)$ given by Eq.~\ref{heurG}, or equivalently, to the elastic modulus $G_c(T)$ given by Eq.~\ref{Gc}, as expected.
	Interestingly, if $\Delta E = 0$, one finds from both above expressions that $w(T) = w^*$ for any temperature $T$, so that Eq.~\ref{wT} gives $G(T) = w^*g_z(p)nk_BT$, which is precisely the elastic modulus of a purely entropic gel.

	As mentioned before, the emergence of the phenomenological factor $w^*$ was not explained in Refs.~\cite{yoshikawa2021prx,sakumichi2021polymj}, so here we suggest that it might be related to a correction on the counting of the number of effective elastic elements in solution.
	In particular, if one considers that the number density of chains between crosslinks can be defined as $n_f = c N_A/M_f$, and that the molar mass $M_f$ of bridged chains between crosslinks is related to the molar mass $M$ of $z$-coordinated precursor molecules as $M_f = 2 (M/z)$, one finds that $n_f = (z/2)n$, where $n$ is given by Eq.~\ref{numberdensity}.
	Hence, if one consistently assumes that the number density of effective elastic elements in solution is, in fact, given by $n_e = g_z(p) n_f \approx w^* g_z(p) n$, so that $n_f \approx w^* n$, these relations lead to
\begin{equation}
w^* \approx \frac{z}{2} ~~,
\end{equation}
which, in the case of tetra-PEG gels where crosslinks have functionality equal to $z=4$, is indeed close to $2.17$.
	In practice, by considering that the above equality is correct, it can be used to recast the function $w(T)$ that is given by Eq.~\ref{wT2} together with its general definition (Eq.~\ref{wT}) in order to yield a final expression for the elastic modulus of our model, that is,
\begin{equation}
G(T) = \left( \frac{ p - 2/z}{1 - 2/z} \right) nk_BT w(T) ~,
\label{elasticmodulusfinal}
\end{equation}
where we recall that $n$ corresponds to the number density of $z$-coordinated precursor molecules (which is also proportional to the number density of crosslinks $n_c$), $p$ is the connectivity of the chains in the network (which should be within the range $2/z \ll p \leq 1$ for the function $g_z(p)$ given by Eq.~\ref{gz} to be valid), and $w(T)$ is the temperature-dependent correction factor, which is found to also depend on $z$ and is given by
\begin{equation}
w(T) = \frac{z}{4} \left(1+e^{-\Delta E/k_BT}\right) \left( \frac{\ell_s + \ell_be^{\Delta E/k_BT}}{\ell_s+\ell_b}\right)~.
\label{wTzdependent}
\end{equation}
	Accordingly, for gels with tetra-functional crosslinks of coordination equal to $z=4$, the term inside the left parenthesis in the above expression will ensures that $w(T)=2$ for any value of $T$ if $\Delta E=0$.
	Even so, it is worth noting that, in general, when $\Delta E=0$, Eq~\ref{wTzdependent} implies that $w(T)=z/2$ for any temperature.

\section{Conclusions}
\label{conclusions}

\vspace{-0.3cm}

	In this study we explored a simple coarse-grained model in order to investigate the origin of the 
negative, $G_E(T)$, and positive, $G_S(T)$, contributions to the elastic modulus $G(T)$ of rubber-like gels.
	By computing an exact expression for the free energy, we were able to find the non-linear temperature-dependent stress-strain relation given by Eq.~\ref{sigma-gamma}, which provided expressions for the positive, $\sigma_S(T,\gamma)$, and the negative, $\sigma_E(T,\gamma)$, contributions to the stress $\sigma(T,\gamma)$ given by Eqs.~\ref{sigmaS} and~\ref{sigmaE}, respectively.

	The theoretical expressions obtained for the elastic moduli $G(T)$, $G_S(T)$, and $G_E(T)$, given by Eqs.~\ref{elasticmodulus-total2},~\ref{GSofT}, and~\ref{GEofT3}, respectively, strongly support the idea that the negative contribution to the elastic modulus of rubber-like gels arises from the effective interaction energy between the chains in the gel network and their neighboring solvent molecules, which is characterized in the present model by a positive energy difference $\Delta E$ and a negative value for $\ell_b$.
	Indeed, our theoretical results indicate that $G_E(T)<0$ if $\Delta E >0$, $\ell_s > 0$, and $\ell_b<0$, while the condition $\Delta E = 0$ implies that there will be no negative ({\it i.e.}, energy-related) contribution to the elastic modulus, that is, $G_S(T) = G(T)$, and thus $G_E(T) = G(T) - G_S(T) = 0$.
	This last conclusion can be made directly from the equations deduced from our model, \textit{i.e.}, Eqs.~\ref{GEofT3} and \ref{sigmaE}, which yields $G_E(T) = 0$ and $\sigma_E(\gamma, T) = 0$ when $\Delta E=0$ for any values of temperature $T$ and strain $\gamma$.
	Also, from Eqs.~\ref{T0temp} and \ref{T0-dE-ls-lb}, one finds that having $T^*_0 = 0$ (or, similarly, $T_0 = 0$) is an equivalent way of concluding that no negative energy-related contribution is present in the system, {\it i.e.}, $\Delta E = 0$.

	Finally, it is worth mentioning that, by comparing our theoretical results and the results obtained from experiments, we show that, although simple, the present model yields expressions that display a good agreement with the experimental data of Ref.~\cite{yoshikawa2021prx} obtained for tetra-PEG gels.
	In particular, we validate our approach by considering the experimental data obtained for $\sigma(T,\gamma)$ at different strains (Fig.~\ref{fig:stress-temp}), and its positive ($\sigma_S$) and negative ($\sigma_E$) contributions as functions of both the strain (Fig.~\ref{fig:stress-gamma}) and the temperature (Fig.~\ref{fig:stresses-temp}).
	In addition, we inferred from the experimental results presented in Fig.~\ref{fig:temp0Xconc_temp0Xmol} that the general expression of the elastic modulus should be in the form of Eq.~\ref{theorG}, {\it i.e.}, $G(T) = g_z(p) n k_BT w(T)$, with $w(T)$ being a temperature-dependent correction factor, $n$ the number density of $z$-coordinated precursor molecules in solution, and $g_z(p)$ a connectivity-dependent function given by Eq.~\ref{gz}.
	Interestingly, the final expression for $G(T)$ obtained from our model, {\it i.e.}, Eq.~\ref{elasticmodulusfinal}, should allow one to describe the mechanical response of both rubber materials ($\Delta E = 0$) and rubber-like gels ($\Delta E > 0$).
	Therefore, we believe that the general discussions presented here may provide a useful theoretical framework which could be used in future studies that want to probe and interpret the elastic modulus of rubber-like gels taking into account its negative energy-related contribution, including those based on numerical simulations~\cite{toda2018aipadv,sugimura2013polymerj,rizzi2016jcp}.

\vspace{0.3cm}

\noindent
{\small {\bf Acknowledgements.} The authors acknowledge the Brazilian agencies FAPEMIG (Process APQ-02783-18), CNPq (N$^{\circ}$ 312999/2021-6 and 426570/2018-9), and CAPES (code 001).}






\appendix

\section{Force-extension relationship in the 
constant-strain ensemble}
\label{constant-strain}


	In this Appendix we indicate how the force-extension relation given by Eq.~\ref{force-dist} can be also obtained from the free energy  $\mathcal{F}(N,T,\ell)$ evaluated in the {\it constant-strain ensemble}.
	First, by considering that $n_b=N-n_s$, one finds that Eq.~\ref{end-to-end-nsnb} yields $\ell= (N-n_s)\ell_b + n_s \ell_s$, which should be within the range $0 \leq \ell < N \ell_s$, hence
\begin{equation}
n_s(\ell) = \frac{\ell - N \ell_b}{\ell_s - \ell_b} 
~~~~\text{and}~~~~
n_b(\ell) = \frac{N \ell_s - \ell}{\ell_s - \ell_b}~~.
\end{equation}
	The above expressions can be used to obtain both the internal energy $U(N,\ell)$ and the entropy $S(N,\ell)$ that are used to evaluate the free energy $\mathcal{F}(N,T,\ell) = U(N,\ell) - T S(N,\ell)$.
	Thus, from such free energy one can obtain the mean force $f$ ({\it i.e.}, the mean stress $\sigma=f/A$) that should be applied to the chain in order to keep its length equal to $\ell$, that is,
\begin{equation}
f(N,T,\ell) = \left[ \frac{\partial }{\partial \ell} \mathcal{F}(N,T,\ell) \right]_{N,T}
\label{barforce}
\end{equation}
	Since the internal energy is given by Eq.~\ref{internalenergy-nsnb}, one have that the energy-related contribution to the force is simply
\begin{equation}
\left[ \frac{\partial}{\partial \ell} U(N,\ell) \right]_{N,T}
= \left( \frac{ E_s  - E_b  }{ \ell_s - \ell_b } \right) 
= \frac{\Delta E}{\Delta \ell} ~~.
\label{Uderivative}
\end{equation}
	Additionally, the entropy can be estimated as $S(N,\ell) = k_B \ln \mathrm{\Omega}(N,\ell)$, with the density of states, Eq.~\ref{dos-nsnb}, given by
\begin{equation}
\mathrm{\Omega}(N,\ell) = \frac{N!}{
\left[ (\ell - N \ell_b)/(\ell_s - \ell_b) \right]!
\left[ (N \ell_s - \ell)/(\ell_s - \ell_b) \right]!
}~~.
\end{equation}
	Here one can consider the Stirling's approximation, {\it i.e.},
$\ln(n!) \approx n\ln(n) - n$,  which yields the derivative of $S(N,\ell)$ with respect to $\ell$ equals to
\begin{eqnarray}
\left[ \frac{\partial }{\partial \ell} S(N,\ell) \right]_{N,T}
&=&
\frac{k_B}{\Delta \ell}
 \ln\left( \frac{N \ell_s - \ell }{\ell - N \ell_b} \right) 
~~. \label{Sderivative} 
\end{eqnarray}
Hence, by replacing the expressions~\ref{Uderivative} and~\ref{Sderivative} in Eq.~\ref{barforce}, one can verify that the result for the force $f$ is equivalent to the expression~\ref{force-dist} derived from the {\it constant-stress ensemble}.

\section{Stress-strain relationship from the Helmholtz free energy density}
\label{Helmholtzfreenergy}

	We include this Appendix in order to indicate that the stress-strain relationship given by expression~\ref{sigma-gamma} can be also evaluated as
\begin{equation}
\sigma(T,\gamma) = \left[ \frac{\partial }{\partial \gamma}
\left( \frac{\mathcal{F}}{V_0} \right) \right]_{T}~,
\label{sigFree}
\end{equation}
 where $\mathcal{F}/V_0$ is the Helmholtz free energy density of the system, with $\mathcal{F} = \mathcal{G} + f \ell$ and $V_0 = A \ell_0$.
	In particular, one can consider that $\mathcal{G}$ and $f$ are given, respectively, by Eqs.~\ref{Gibbs} and~\ref{force-dist}, replace $\ell$ by $\gamma$ through Eq.~\ref{gamma-strain}, and then use Eq.~\ref{ell0} for $\ell_0$, which yields
\begin{eqnarray}
\dfrac{\mathcal{F}}{V_0}
= &-&(\gamma_b+\gamma_s) \dfrac{k_BT}{A\Delta\ell}\ln\left[\left(\dfrac{~\gamma_b +\gamma_s}{\gamma_b+ \gamma}\right)\left( \dfrac{\gamma_b +\gamma}{\gamma_s - \gamma} \right)^{\ell_s/\Delta\ell}\right]\nonumber\\
&+&\,(1+\gamma)\left[\dfrac{\Delta E}{A\Delta\ell} + \dfrac{k_BT}{A\Delta\ell}\ln\left(\dfrac{\gamma_b + \gamma}{\gamma_s-\gamma}\right) \right]\nonumber\\
&+&(\gamma_b+\gamma_s)\dfrac{\Delta \tilde{E}}{A\Delta\ell}~~,
\label{FreeEnergy}
\end{eqnarray}
where $\Delta \tilde{E} = (\ell_sE_b - \ell_bE_s)/\Delta\ell$.
	It is worth emphasizing that,
as we have mentioned in Sec.~\ref{Tgamma-depend-sigma}, the response of the whole rubber-like material is obtained by analyzing the properties of its fundamental building block, {\it i.e.}, a single bridged chain immersed in a incompressible fluid solvent.
	Arguably, since there is not a preferable orientation for the chains and the material is expected to be disordered and isotropic at mesoscopic length scales, we assume that this approach leads to the same result as if we had analyzed the whole system and the stress depends only on a scalar Helmholtz free energy density which is given by Eq.~\ref{FreeEnergy}.

\section{Definitions of $f_S$ and $f_E$}
\label{definitions-fe-fs}
	In this Appendix we explicitly discuss the alternative definitions of the $S$ and $E$ contributions to the stress that we and the authors of Ref.~\cite{yoshikawa2021prx} explored, since it seems to differ from the usual definitions adopted in the literature~\cite{flory1953book}.
	In general, for isotropic rubber-like materials, one usually considers the following Helmholtz free energy
	\begin{equation}
	\mathcal{F}(V,T,\ell)= U(V,T,\ell) - TS(V,T,\ell)~,
	\label{FreeH}
	\end{equation}
which differential form is given by
	\begin{equation}
	d\mathcal{F}=-PdV - SdT + fd\ell~,
	\label{diffFreeEnergy}
	\end{equation}
where $V$, $T$, and $\ell$ are the volume, the temperature, and the axial length of the material,
while $P$, $S$, and $f$ are the hydrostatic pressure, the entropy, and the axial load, 
respectively.
	Hence, from Eq.~\ref{diffFreeEnergy}, one can write the equation of state of such material as 
	\begin{equation}
	f(V,T,\ell)=\left[\dfrac{\partial }{\partial \ell}\mathcal{F}(V,T,\ell)\right]_{T,V}~,
	\label{EoE0}
	\end{equation}
so that Eq.~\ref{FreeH} leads to
	\begin{equation}
	f(V,T,\ell) = \left[\dfrac{\partial }{\partial \ell}U(V,T,\ell)\right]_{T,V} - T\left[\dfrac{\partial }{\partial \ell}S(V,T,\ell)\right]_{T,V}~.
	\label{EoE1}
	\end{equation}
	Usually, in order to determine the change in the entropy in the second term at the right side of the above equation, one could think of using the Maxwell relation~\cite{yoshikawa2021prx} $-(\partial S/\partial T)_{T,V} = (\partial f/\partial T)_{T,\ell}$. 
	As it happens, the experimental setup to measure the derivative $(\partial f/\partial T)_{V,\ell}$ is
not very practical,
so to overcome this issue one is led to the following approximation~\cite{flory1953book}
	\begin{equation}
	-\left[\dfrac{\partial }{\partial \ell}S(V,T,\ell)\right]_{T,V}\approx \left[\dfrac{\partial }{\partial T}f(V,T,\ell)\right]_{P,\lambda}~,
	\end{equation}
which turns Eq.~\ref{EoE1} into
	\begin{equation}
	f(V,T,\ell) \approx \left[\dfrac{\partial }{\partial \ell}U(V,T,\ell)\right]_{T,V} + T\left[\dfrac{\partial }{\partial T}f(V,T,\ell)\right]_{P,\lambda}~,
	\label{approx}
	\end{equation}
	where $\lambda= \ell/\ell_0=1+\gamma$ is the elongation of the material, and $\gamma$ is the strain defined as in Eq.~\ref{gamma-strain}.
	It is worth metioning that, usually, the definitions of the energetic and the entropic contributions to the force are given, respectively, by the first and the second terms in the above expression~\cite{flory1953book}.

	Interestingly, although the experiments in Ref.~\cite{yoshikawa2021prx} were carried out in a fixed strain setup and at a constant pressure, the subscripts $\gamma$ and $P$ were suppressed in the notation of their thermodynamic derivatives.
	Even so, since we have decided to put forward a modelling approach aiming to describe their experimental data, we avoid approximations by considering that, since $\gamma$ is defined through Eq.~\ref{gamma-strain}, 
so $\lambda$ must also depend on the temperature $T$.
	In that scenario the differential form of $\ell$ should be written as
\begin{equation}
d\ell = \left( \frac{\partial \ell}{\partial \lambda} \right)_{T} \, d\lambda + \left( \frac{ \partial \ell}{\partial T} \right)_\lambda dT~. 
\end{equation}
	Thus, through Eq.~\ref{diffFreeEnergy}, and recalling the fact that the PEG hydrogel can be considered as an incompressible fluid~\cite{yoshikawa2021prx}, \textit{i.e.,} $dV = 0$, one may rewrite the differential form of the free energy  
as
	\begin{equation}
	d\mathcal{F}=\left[ -S + f\left(\dfrac{\partial \ell }{\partial T}\right)_{\lambda}\right] dT + 
f \left( \frac{\partial \ell}{\partial \lambda} \right)_{T}\, d\lambda~,
	\label{FreeH-lambda}
	\end{equation}
so that
	\begin{equation}
	\left[\dfrac{\partial }{\partial T}\mathcal{F}(T,\ell)\right]_{\lambda} = -S(T,\ell) + f(T,\ell)\left[\dfrac{\partial \ell(T,\lambda)}{\partial T}\right]_{\lambda}~.
\label{partF-partT}
	\end{equation}
	Now, from Eq.~\ref{EoE0}, one has that
	\begin{equation}
	\left[\dfrac{\partial }{\partial T}f(T,\ell)\right]_{\lambda} = \left[\dfrac{\partial}{\partial T}\left[\dfrac{\partial }{\partial \ell}\mathcal{F}(T,\ell)\right]_{T}\right]_{\lambda} ~, 
	\end{equation}
thus, through the
identity between second derivatives
and by considering the result given by Eq.~\ref{partF-partT}, one finds that
	\begin{equation}
	\left[\dfrac{\partial }{\partial T}f(T,\ell)\right]_{\lambda}
	= \left[\dfrac{\partial}{\partial \ell}\left(-S +  f\left(\dfrac{\partial \ell}{\partial T}\right)_{\lambda}\right) \right]_{T} ~.
	\label{force-temp_lambda}
	\end{equation}
	Accordingly, in this work we assume that the positive ($S$) contribution to the force is defined as
	\begin{equation}
	f_S(T,\ell) = T\left[\dfrac{\partial }{\partial T}f(T,\ell)\right]_\lambda~,
	\end{equation}
with the derivative of $f$ given by Eq.~\ref{force-temp_lambda}, that is,
	\begin{equation}
	f_S(T,\ell) = 
- T \left[ 
\frac{\partial}{\partial \ell}
\left[
S(T,\ell) - f(T,\ell)
\left[
\frac{\partial}{\partial T}
\ell (T,\lambda)
\right]_{\lambda}
\right]
\right]_{T}~.
	\label{fs}
	\end{equation}
	Furthermore, with that definition, the energy-related ($E$) contribution to the force can be evaluated as $f_{E}(T,\ell) = f(T,\ell) - f_{S}(T,\ell)$, which yields
	\begin{equation}
	f_E(T,\ell) = \left[\dfrac{\partial}{\partial \ell}\left[U(T,\ell) - T f(T,\ell)  \left[\dfrac{\partial }{\partial T} \ell(T,\lambda)\right]_{\lambda}\right]\right]_{T}~.
	\label{fe}
	\end{equation}
	Hence, in contrast to the usual definitions that have been adopted in the literature~\cite{flory1953book,yasuda2009macromo},  one have that $f_S \ne -T(\partial S/\partial \ell)_{T,V}$ and $f_E \ne (\partial U/\partial \ell)_{T,V}$.
	Even though neither Eq.~\ref{fs} nor Eq.~\ref{fe} correspond to the usual expressions
for the energetic and entropic contributions, respectively, the sum of the two expressions lead to the very same equation of state, {\it i.e.}, Eq.~\ref{EoE1}.
	Also, from the above results for $f_S(T,\ell)$ and $f_E(T,\ell)$, and by considering $\ell =\lambda\ell_0=(1+\gamma)\ell_0$ (see Eq.~\ref{gamma-strain}), one have that the expressions for $\sigma_S(T,\gamma)$ and $\sigma_E(T,\gamma)$ can be evaluated as
	\begin{equation}
	\sigma_S(T,\gamma)  =\dfrac{1}{A}f_S(T,\ell)|_{\ell=(1+\gamma)\ell_0}^{\,}
	\end{equation}
and
	\begin{equation}
	\sigma_E(T,\gamma)  =\dfrac{1}{A}f_E(T,\ell)|_{\ell=(1+\gamma)\ell_0}^{\,}~.
	\end{equation}
	It can be easily verified that these definitions are not only fully consistent with the expressions derived in Sec.~\ref{S-and-E-contrib}, but also that they are also consistent with the way that the experimental results were obtained in Ref.~\cite{yoshikawa2021prx}.



\begin{thebibliography}{33}
\ifx \bisbn   \undefined \def \bisbn  #1{ISBN #1}\fi
\ifx \binits  \undefined \def \binits#1{#1}\fi
\ifx \bauthor  \undefined \def \bauthor#1{#1}\fi
\ifx \batitle  \undefined \def \batitle#1{#1}\fi
\ifx \bjtitle  \undefined \def \bjtitle#1{#1}\fi
\ifx \bvolume  \undefined \def \bvolume#1{\textbf{#1}}\fi
\ifx \byear  \undefined \def \byear#1{#1}\fi
\ifx \bissue  \undefined \def \bissue#1{#1}\fi
\ifx \bfpage  \undefined \def \bfpage#1{#1}\fi
\ifx \blpage  \undefined \def \blpage #1{#1}\fi
\ifx \burl  \undefined \def \burl#1{\textsf{#1}}\fi
\ifx \doiurl  \undefined \def \doiurl#1{\url{https://doi.org/#1}}\fi
\ifx \betal  \undefined \def \betal{\textit{et al.}}\fi
\ifx \binstitute  \undefined \def \binstitute#1{#1}\fi
\ifx \binstitutionaled  \undefined \def \binstitutionaled#1{#1}\fi
\ifx \bctitle  \undefined \def \bctitle#1{#1}\fi
\ifx \beditor  \undefined \def \beditor#1{#1}\fi
\ifx \bpublisher  \undefined \def \bpublisher#1{#1}\fi
\ifx \bbtitle  \undefined \def \bbtitle#1{#1}\fi
\ifx \bedition  \undefined \def \bedition#1{#1}\fi
\ifx \bseriesno  \undefined \def \bseriesno#1{#1}\fi
\ifx \blocation  \undefined \def \blocation#1{#1}\fi
\ifx \bsertitle  \undefined \def \bsertitle#1{#1}\fi
\ifx \bsnm \undefined \def \bsnm#1{#1}\fi
\ifx \bsuffix \undefined \def \bsuffix#1{#1}\fi
\ifx \bparticle \undefined \def \bparticle#1{#1}\fi
\ifx \barticle \undefined \def \barticle#1{#1}\fi
\ifx \bconfdate \undefined \def \bconfdate #1{#1}\fi
\ifx \botherref \undefined \def \botherref #1{#1}\fi
\ifx \url \undefined \def \url#1{\textsf{#1}}\fi
\ifx \bchapter \undefined \def \bchapter#1{#1}\fi
\ifx \bbook \undefined \def \bbook#1{#1}\fi
\ifx \bcomment \undefined \def \bcomment#1{#1}\fi
\ifx \oauthor \undefined \def \oauthor#1{#1}\fi
\ifx \citeauthoryear \undefined \def \citeauthoryear#1{#1}\fi
\ifx \endbibitem  \undefined \def \endbibitem {}\fi
\ifx \bconflocation  \undefined \def \bconflocation#1{#1}\fi
\ifx \arxivurl  \undefined \def \arxivurl#1{\textsf{#1}}\fi
\csname PreBibitemsHook\endcsname

\bibitem{richtering2014softmatter}
\begin{barticle}
\bauthor{\bsnm{Richtering}, \binits{W.}},
\bauthor{\bsnm{Saunders}, \binits{B.R.}}:
\batitle{Gel architectures and their complexity}.
\bjtitle{Soft Matter}
\bvolume{10},
\bfpage{3695}
(\byear{2014})
\end{barticle}
\endbibitem

\bibitem{annabi2014advmater}
\begin{barticle}
\bauthor{\bsnm{Annabi}, \binits{N.}},
\bauthor{\bsnm{Tamayol}, \binits{A.}},
\bauthor{\bsnm{Uquillas}, \binits{J.A.}},
\bauthor{\bsnm{Akbari}, \binits{M.}},
\bauthor{\bsnm{Bertassoni}, \binits{L.E.}},
\bauthor{\bsnm{Cha}, \binits{C.}},
\bauthor{\bsnm{Camci-Unal}, \binits{G.}},
\bauthor{\bsnm{Dokmeci}, \binits{M.R.}},
\bauthor{\bsnm{Peppas}, \binits{N.A.}},
\bauthor{\bsnm{Khademhosseini}, \binits{A.}}:
\batitle{Rational design and applications of hydrogels in regenerative
  medicine}.
\bjtitle{Adv. Mater.}
\bvolume{26},
\bfpage{85}
(\byear{2014})
\end{barticle}
\endbibitem

\bibitem{thiele2014advmater}
\begin{barticle}
\bauthor{\bsnm{Thiele}, \binits{J.}},
\bauthor{\bsnm{Ma}, \binits{Y.}},
\bauthor{\bsnm{Bruekers}, \binits{S.M.C.}},
\bauthor{\bsnm{Ma}, \binits{S.}},
\bauthor{\bsnm{Huck}, \binits{W.T.S.}}:
\batitle{Designer hydrogels for cell cultures: A materials selection guide}.
\bjtitle{Adv. Mater.}
\bvolume{26},
\bfpage{125}
(\byear{2014})
\end{barticle}
\endbibitem

\bibitem{kamata2015advhmater}
\begin{barticle}
\bauthor{\bsnm{Kamata}, \binits{H.}},
\bauthor{\bsnm{Li}, \binits{X.}},
\bauthor{\bsnm{Chung}, \binits{U.-i.}},
\bauthor{\bsnm{Sakai}, \binits{T.}}:
\batitle{Design of hydrogels for biomedical applications}.
\bjtitle{Adv. Healthcare Mater}
\bvolume{4},
\bfpage{2360}
(\byear{2015})
\end{barticle}
\endbibitem

\bibitem{calo2015eurpolj}
\begin{barticle}
\bauthor{\bsnm{Cal\'o}, \binits{E.}},
\bauthor{\bsnm{Khutoryanskiy}, \binits{V.V.}}:
\batitle{Biomedical applications of hydrogels: A review of patents and
  commercial products}.
\bjtitle{Eur. Polym. J.}
\bvolume{65},
\bfpage{252}
(\byear{2015})
\end{barticle}
\endbibitem

\bibitem{basu2011macromol}
\begin{barticle}
\bauthor{\bsnm{Basu}, \binits{A.}},
\bauthor{\bsnm{Wen}, \binits{Q.}},
\bauthor{\bsnm{Mao}, \binits{X.}},
\bauthor{\bsnm{Lubensky}, \binits{T.C.}},
\bauthor{\bsnm{Janmey}, \binits{P.A.}},
\bauthor{\bsnm{Yodh}, \binits{A.G.}}:
\batitle{Nonaffine displacements in flexible polymer networks}.
\bjtitle{Macromolecules}
\bvolume{44},
\bfpage{1671}
(\byear{2011})
\end{barticle}
\endbibitem

\bibitem{wen2012softmatter}
\begin{barticle}
\bauthor{\bsnm{Wen}, \binits{Q.}},
\bauthor{\bsnm{Basu}, \binits{A.}},
\bauthor{\bsnm{Janmey}, \binits{P.A.}},
\bauthor{\bsnm{Yodh}, \binits{A.G.}}:
\batitle{Non-affine deformations in polymer hydrogels}.
\bjtitle{Soft Matter}
\bvolume{8},
\bfpage{8039}
(\byear{2012})
\end{barticle}
\endbibitem

\bibitem{dai2021softmatter}
\begin{barticle}
\bauthor{\bsnm{Dai}, \binits{Y.}},
\bauthor{\bsnm{Zhang}, \binits{R.}},
\bauthor{\bsnm{Sun}, \binits{W.}},
\bauthor{\bsnm{Wang}, \binits{T.}},
\bauthor{\bsnm{Chen}, \binits{Y.}},
\bauthor{\bsnm{Tong}, \binits{Z.}}:
\batitle{Dynamical heterogeneity in the gelation process of a polymer solution
  with a lower critical solution temperature}.
\bjtitle{Soft Matter}
\bvolume{17},
\bfpage{3222}
(\byear{2021})
\end{barticle}
\endbibitem

\bibitem{gu2020angewchem}
\begin{barticle}
\bauthor{\bsnm{Gu}, \binits{Y.}},
\bauthor{\bsnm{Zhao}, \binits{J.}},
\bauthor{\bsnm{Johnson}, \binits{J.A.}}:
\batitle{A unifying review of polymer networks: From rubbers and gels to porous
  frameworks}.
\bjtitle{Angew. Chem. Int.}
\bvolume{59},
\bfpage{5022}--\blpage{5049}
(\byear{2020})
\end{barticle}
\endbibitem

\bibitem{danielsen2021chemrev}
\begin{barticle}
\bauthor{\bsnm{Danielsen}, \binits{S.P.O.}},
\bauthor{\bsnm{Beech}, \binits{H.K.}},
\bauthor{\bsnm{Wang}, \binits{S.}},
\bauthor{\bsnm{El-Zaatari}, \binits{B.M.}},
\bauthor{\bsnm{Wang}, \binits{X.}},
\bauthor{\bsnm{Sapir}, \binits{L.}},
\bauthor{\bsnm{Ouchi}, \binits{T.}},
\bauthor{\bsnm{Wang}, \binits{Z.}},
\bauthor{\bsnm{Johnson}, \binits{P.N.}},
\bauthor{\bsnm{Hu}, \binits{Y.}},
\bauthor{\bsnm{Lundberg}, \binits{D.J.}},
\bauthor{\bsnm{Stoychev}, \binits{G.}},
\bauthor{\bsnm{Craig}, \binits{S.L.}},
\bauthor{\bsnm{Johnson}, \binits{J.A.}},
\bauthor{\bsnm{Kalow}, \binits{J.A.}},
\bauthor{\bsnm{Olsen}, \binits{B.D.}},
\bauthor{\bsnm{Rubinstein}, \binits{M.}}:
\batitle{Molecular characterization of polymer networks}.
\bjtitle{Chem. Rev.}
\bvolume{121},
\bfpage{5042}--\blpage{5092}
(\byear{2021})
\end{barticle}
\endbibitem

\bibitem{yoshikawa2021prx}
\begin{barticle}
\bauthor{\bsnm{Yoshikawa}, \binits{Y.}},
\bauthor{\bsnm{Sakumichi}, \binits{N.}},
\bauthor{\bsnm{Chung}, \binits{U.-i.}},
\bauthor{\bsnm{Sakai}, \binits{T.}}:
\batitle{Negative energy elasticity in a rubberlike gel}.
\bjtitle{Phys. Rev. X}
\bvolume{11},
\bfpage{011045}
(\byear{2021})
\end{barticle}
\endbibitem

\bibitem{treloarbook}
\begin{bbook}
\bauthor{\bsnm{Treloar}, \binits{L.R.G.}}:
\bbtitle{The Physics of Rubber Elasticity}.
\bpublisher{Oxford University Press},
\blocation{Oxford}
(\byear{1975})
\end{bbook}
\endbibitem

\bibitem{toda2018aipadv}
\begin{barticle}
\bauthor{\bsnm{Toda}, \binits{M.}},
\bauthor{\bsnm{Morita}, \binits{H.}}:
\batitle{Rubber elasticity of realizable ideal networks}.
\bjtitle{AIP Advances}
\bvolume{8},
\bfpage{125005}
(\byear{2018})
\end{barticle}
\endbibitem

\bibitem{sakumichi2021polymj}
\begin{barticle}
\bauthor{\bsnm{Sakumichi}, \binits{N.}},
\bauthor{\bsnm{Yoshikawa}, \binits{Y.}},
\bauthor{\bsnm{Sakai}, \binits{T.}}:
\batitle{Linear elasticity of polymer gels in terms of negative energy
  elasticity}.
\bjtitle{Polym. J.}
\bvolume{53},
\bfpage{1293}
(\byear{2021})
\end{barticle}
\endbibitem

\bibitem{flory1953book}
\begin{bbook}
\bauthor{\bsnm{Flory}, \binits{P.J.}}:
\bbtitle{Principles of Polymer Chemistry}.
\bpublisher{Cornell University Press},
\blocation{Ithaca}
(\byear{1953})
\end{bbook}
\endbibitem

\bibitem{james1953jcp}
\begin{barticle}
\bauthor{\bsnm{James}, \binits{H.M.}},
\bauthor{\bsnm{Guth}, \binits{E.}}:
\batitle{Statistical thermodynamics of rubber elasticity}.
\bjtitle{J. Chem. Phys.}
\bvolume{21},
\bfpage{1039}--\blpage{1049}
(\byear{1953})
\end{barticle}
\endbibitem

\bibitem{flory1977jcp}
\begin{barticle}
\bauthor{\bsnm{Flory}, \binits{P.J.}}:
\batitle{Theory of elasticity of polymer networks. \uppercase{T}he effect of
  local constraints on junctions}.
\bjtitle{J. Chem. Phys.}
\bvolume{66},
\bfpage{5720}--\blpage{5729}
(\byear{1977})
\end{barticle}
\endbibitem

\bibitem{anthony1942jpc}
\begin{barticle}
\bauthor{\bsnm{Anthony}, \binits{R.L.}},
\bauthor{\bsnm{Caston}, \binits{R.H.}},
\bauthor{\bsnm{Guth}, \binits{E.}}:
\batitle{Equations of state for natural and synthetic rubber-like materials.
  \uppercase{I}. \uppercase{U}naccelerated natural soft rubber}.
\bjtitle{J. Chem. Phys.}
\bvolume{46},
\bfpage{826}
(\byear{1942})
\end{barticle}
\endbibitem

\bibitem{fujiyabu2021prl}
\begin{barticle}
\bauthor{\bsnm{Fujiyabu}, \binits{T.}},
\bauthor{\bsnm{Sakai}, \binits{T.}},
\bauthor{\bsnm{Kudo}, \binits{R.}},
\bauthor{\bsnm{Yoshikawa}, \binits{Y.}},
\bauthor{\bsnm{Katashima}, \binits{T.}},
\bauthor{\bsnm{Chung}, \binits{U.I.}},
\bauthor{\bsnm{Sakumichi}, \binits{N.}}:
\batitle{Temperature dependence of polymer network diffusion}.
\bjtitle{Phys. Rev. Lett.}
\bvolume{127},
\bfpage{237801}
(\byear{2021})
\end{barticle}
\endbibitem

\bibitem{shirai2022arxiv}
\begin{barticle}
\bauthor{\bsnm{Shirai}, \binits{N.C.}},
\bauthor{\bsnm{Sakumichi}, \binits{N.}}:
\batitle{Negative energetic elasticity of lattice polymer chain in solvent}.
\bjtitle{arXiv:}
\bvolume{2202.12483},
\bfpage{1}--\blpage{6}
(\byear{2022});
\end{barticle}
\begin{barticle}
\batitle{Solvent-Induced Negative Energetic Elasticity in a Lattice Polymer Chain}.
\bjtitle{Phys. Rev. Lett.}
\bvolume{130},
\bfpage{148101}
(\byear{2023})
\end{barticle}
\endbibitem

\bibitem{junghans2006prl}
\begin{barticle}
\bauthor{\bsnm{Junghans}, \binits{C.}},
\bauthor{\bsnm{Bachmann}, \binits{M.}},
\bauthor{\bsnm{Janke}, \binits{W.}}:
\batitle{Microcanonical analyses of peptide aggregation processes}.
\bjtitle{Phys. Rev. Lett.}
\bvolume{97},
\bfpage{218103}
(\byear{2006})
\end{barticle}
\endbibitem

\bibitem{chen2008pre}
\begin{barticle}
\bauthor{\bsnm{Chen}, \binits{T.}},
\bauthor{\bsnm{Lin}, \binits{X.}},
\bauthor{\bsnm{Liu}, \binits{Y.}},
\bauthor{\bsnm{Lu}, \binits{T.}},
\bauthor{\bsnm{Liang}, \binits{H.}}:
\batitle{Microcanonical analyses of homopolymer aggregation processes}.
\bjtitle{Phys. Rev. E}
\bvolume{78},
\bfpage{056101}
(\byear{2008})
\end{barticle}
\endbibitem

\bibitem{liu2012jcp}
\begin{barticle}
\bauthor{\bsnm{Liu}, \binits{Y.}},
\bauthor{\bsnm{Kellogg}, \binits{E.}},
\bauthor{\bsnm{Liang}, \binits{H.}}:
\batitle{Canonical and micro-canonical analysis of folding of trpzip2: An
  all-atom replica exchange \uppercase{M}onte \uppercase{C}arlo simulation
  study}.
\bjtitle{J. Chem. Phys.}
\bvolume{137},
\bfpage{045103}
(\byear{2012})
\end{barticle}
\endbibitem

\bibitem{frigori2013jcp}
\begin{barticle}
\bauthor{\bsnm{Frigori}, \binits{R.B.}},
\bauthor{\bsnm{Rizzi}, \binits{L.G.}},
\bauthor{\bsnm{Alves}, \binits{N.A.}}:
\batitle{Microcanonical thermostatistics of coarse-grained proteins with
  amyloidogenic propensity}.
\bjtitle{J. Chem. Phys.}
\bvolume{138},
\bfpage{015102}
(\byear{2013})
\end{barticle}
\endbibitem

\bibitem{kubobook}
\begin{bbook}
\bauthor{\bsnm{Kubo}, \binits{R.}}:
\bbtitle{Statistical Mechanics: An Advanced Course with Problems and
  Solutions}.
\bpublisher{North-Holland Physics},
\blocation{New York}
(\byear{1988})
\end{bbook}
\endbibitem

\bibitem{nishi2015jcp}
\begin{barticle}
\bauthor{\bsnm{Nishi}, \binits{K.}},
\bauthor{\bsnm{Noguchi}, \binits{H.}},
\bauthor{\bsnm{Sakai}, \binits{T.}},
\bauthor{\bsnm{Shibayama}, \binits{M.}}:
\batitle{Rubber elasticity for percolation network consisting of
  \uppercase{G}aussian chains}.
\bjtitle{J. Chem. Phys.}
\bvolume{143},
\bfpage{184905}
(\byear{2015})
\end{barticle}
\endbibitem

\bibitem{rubinsteinbook}
\begin{bbook}
\bauthor{\bsnm{Rubinstein}, \binits{M.}},
\bauthor{\bsnm{Colby}, \binits{R.H.}}:
\bbtitle{Polymer Physics}.
\bpublisher{Oxford University Press},
\blocation{Pennsylvania}
(\byear{2003})
\end{bbook}
\endbibitem

\bibitem{reith2003jcomputchem}
\begin{barticle}
\bauthor{\bsnm{Reith}, \binits{D.}},
\bauthor{\bsnm{Mathias~P\"utz}, \binits{F.M.-P.}}:
\batitle{Deriving effective mesoscale potentials from atomistic simulations}.
\bjtitle{J. Comput. Chem.}
\bvolume{13},
\bfpage{1624}
(\byear{2003})
\end{barticle}
\endbibitem

\bibitem{sakai2008macromo}
\begin{barticle}
\bauthor{\bsnm{Sakai}, \binits{T.}},
\bauthor{\bsnm{Matsunaga}, \binits{T.}},
\bauthor{\bsnm{Yamamoto}, \binits{Y.}},
\bauthor{\bsnm{Ito}, \binits{C.}},
\bauthor{\bsnm{Yoshida}, \binits{R.}},
\bauthor{\bsnm{Suzuki}, \binits{S.}},
\bauthor{\bsnm{Sasaki}, \binits{N.}},
\bauthor{\bsnm{Shibayama}, \binits{M.}},
\bauthor{\bsnm{Chung}, \binits{U.I.}}:
\batitle{Design and fabrication of a high-strength hydrogel with ideally
  homogeneous network structure from tetrahedron-like}.
\bjtitle{Macromolecules}
\bvolume{41},
\bfpage{5379}
(\byear{2008})
\end{barticle}
\endbibitem

\bibitem{yasuda2009macromo}
\begin{barticle}
\bauthor{\bsnm{Matsunaga}, \binits{T.}},
\bauthor{\bsnm{Sakai}, \binits{T.}},
\bauthor{\bsnm{Akagi}, \binits{Y.}},
\bauthor{\bsnm{Chung}, \binits{U.I.}},
\bauthor{\bsnm{Shibayama}, \binits{M.}}:
\batitle{\uppercase{SANS} and \uppercase{SLS} studies on tetra-arm
  \uppercase{PEG} gels in as-prepared and swollen states}.
\bjtitle{Macromolecules}
\bvolume{42},
\bfpage{6245}
(\byear{2009})
\end{barticle}
\endbibitem

\bibitem{kolberg2019JACS}
\begin{barticle}
\bauthor{\bsnm{Kolberg}, \binits{A.}},
\bauthor{\bsnm{Wenzel}, \binits{C.}},
\bauthor{\bsnm{Hackenstrass}, \binits{K.}},
\bauthor{\bsnm{Schwarzl}, \binits{R.}},
\bauthor{\bsnm{R\"uttiger}, \binits{C.}},
\bauthor{\bsnm{Hugel}, \binits{T.}},
\bauthor{\bsnm{Gallei}, \binits{M.}},
\bauthor{\bsnm{Netz}, \binits{R.R.}},
\bauthor{\bsnm{Balzer}, \binits{B.N.}}:
\batitle{Opposing temperature dependence of the stretching response of single
  \uppercase{PEG} and \uppercase{PN}i\uppercase{PAM} polymers}.
\bjtitle{J. Am. Chem. Soc.}
\bvolume{141},
\bfpage{11603}--\blpage{11613}
(\byear{2019})
\end{barticle}
\endbibitem

\bibitem{sugimura2013polymerj}
\begin{barticle}
\bauthor{\bsnm{Sugimura}, \binits{A.}},
\bauthor{\bsnm{Asai}, \binits{M.}},
\bauthor{\bsnm{Matsunaga}, \binits{T.}},
\bauthor{\bsnm{Akagi}, \binits{Y.}},
\bauthor{\bsnm{Sakai}, \binits{T.}},
\bauthor{\bsnm{Noguchi}, \binits{H.}},
\bauthor{\bsnm{Shibayama}, \binits{M.}}:
\batitle{Mechanical properties of a polymer network of tetra-\uppercase{PEG}
  gel}.
\bjtitle{Polymer J.}
\bvolume{45},
\bfpage{300}
(\byear{2013})
\end{barticle}
\endbibitem

\bibitem{rizzi2016jcp}
\begin{barticle}
\bauthor{\bsnm{Rizzi}, \binits{L.G.}},
\bauthor{\bsnm{Levin}, \binits{Y.}}:
\batitle{Influence of network topology on the swelling of polyelectrolyte
  nanogels}.
\bjtitle{J. Chem. Phys.}
\bvolume{144},
\bfpage{114903}
(\byear{2016})
\end{barticle}
\endbibitem

\end{thebibliography}
%
%
%


\end{document}